\documentclass[mathpazo]{cicp}

\usepackage{mathrsfs}
\usepackage{subcaption}
\usepackage{amssymb}
\usepackage{amsbsy}
\usepackage[mathscr]{euscript}
\begin{document}

\title[short title]{Time Symmetric Methods}
\title{Conservative Evolution of Black Hole Perturbations with Time-Symmetric Numerical Methods}

\author[O'Boyle, M. F. et.~al.]{
Michael F. O'Boyle\affil{1},
Charalampos Markakis\affil{2,3,4}\comma\corrauth, 
Lidia J. Gomes Da Silva\affil{2},
Rodrigo Panosso Macedo\affil{5},
Juan A. Valiente Kroon\affil{2}
}
\address{
    \affilnum{1}\ Department of Physics, University of Illinois at Urbana-Champaign, Urbana, Illinois 61801, USA,\\
    \affilnum{2}\ School of Mathematical Sciences, Queen Mary University of London, E1 4NS, London, UK\\
    \affilnum{3}\ DAMTP, Centre for
Mathematical Sciences, University of Cambridge, CB3 0WA, Cambridge, UK\\
\affilnum{4}\ NCSA, University of Illinois at Urbana-Champaign, Urbana, Illinois 61801, USA\\
    \affilnum{5}\ Mathematical Sciences, University of Southampton, SO17 1BJ, Southampton, UK \\
}
\email{{\tt c.markakis@qmul.ac.uk}}

\begin{abstract}
The scheduled launch of the LISA Mission in the next decade has called attention to the gravitational self-force problem. Despite an extensive body of theoretical work, long-time numerical computations of gravitational waves from extreme-mass-ratio-inspirals remain challenging.  This work proposes a class of numerical evolution schemes suitable to this problem based on Hermite integration. Their most important feature is time-reversal symmetry and unconditional stability, which enables these methods to preserve symplectic structure, energy, momentum and other Noether charges over long time periods. We apply Noether's theorem to the master fields of black hole perturbation theory on a hyperboloidal slice of Schwarzschild spacetime to show that there exist constants of evolution that numerical simulations must preserve. We demonstrate that time-symmetric integration schemes based on a 2-point Taylor expansion (such as Hermite integration) numerically conserve these quantities, unlike schemes based on a 1-point Taylor expansion (such as Runge-Kutta). This makes time-symmetric schemes ideal for long-time EMRI simulations.
\end{abstract}


\maketitle
\section{Introduction}
\label{Intro}
The direct detections of gravitational radiation from compact binary coalescence by the LIGO-Virgo-KAGRA (LVK) Scientific Collaboration in recent years has created a surge of interest in gravitational wave (GW) science. Supplementing electromagnetic and particle observations, we now have an additional window through which to view the universe \cite{the_ligo_scientific_collaboration_gwtc-1:_2018}. The events most likely to create observable GW events involve black holes, owing to their compactness and strong curvature of spacetime \cite{barausse_prospects_2020,moore_gravitational-wave_2015,amaro-seoane_intermediate_2007}. However, the LVK detector frequency band is only sensitive to events where the progenitors have comparable mass \cite{moore_gravitational-wave_2015}, with the largest confirmed mass ratio observed to date being $\sim 1:9$ \cite{LIGOScientific:2020zkf} \footnote{A merger with an estimated mass ratio $\sim 1:26$ was reported, but LVK concedes that this ratio is beyond the capabilities of their models and reported the strong possibility of systematic errors \cite{LIGOScientific:2021djp}.}. 

Another promising channel for observations are extreme-mass-ratio-inspirals (EMRIs) where a star or stellar-mass black hole orbits then plunges into a supermassive black hole. Such events are expected to be regular occurrences in galactic centers and would provide numerous astrophysical insights  \cite{barausse_prospects_2020,amaro-seoane_intermediate_2007}. A major complication in their study is the difficulty entailed in accurately simulating the orbits of the smaller object and computing the GWs emitted.  Standard numerical relativity is poorly equipped to handle this problem: the objects' disparate masses creates two vastly different length scales, requiring a fine grid and small timesteps to accurately resolve, making long-time simulations computationally intractable. A more promising route is the gravitational self-force program, where the smaller object is modeled as a point mass that moves on a stationary background spacetime. It sources linear perturbations that result in radiation reaction and self-force effects \cite{barack_self-force_2019,poisson_motion_2011}. It has been shown that, in a radiation gauge, these effects can be derived by reconstructing the metric from curvature scalars \cite{barack_time-domain_2017,pound_gravitational_2014,keidl_gravitational_2010,Shah:2010bi}. Thus, the accurate evolution of scalar fields in curved spacetime has direct bearing on problems in GW science and relativistic astrophysics.

Numerical relativity studies seem to favor explicit time-evolution schemes, like the classical Runge-Kutta methods. Although they are easy to implement and well-studied, they suffer from two drawbacks: they are conditionally stable, that is, CFL limited, and known to violate energy conservation and symplectic structure in Hamiltonian systems. In GW computations, it is vital to accurately track the energy a system loses to radiation, and, with an explicit scheme, it is unclear \textit{a priori} whether energy loss is due to radiative loss or truncation error or other numerical dissipation.  A preferable alternative is a so-called \textit{geometric integrator} which respects a qualitative feature of Hamiltonian dynamics, like symplecticity or time-reversal symmetry \cite{hairer_geometric_2006,sanz-serna_numerical_1994}. Time-symmetry is a particularly appealing property, since Noether's theorem relates time-translation symmetry to energy conservation. Moreover, such geometric methods often possess enhanced stability properties. Geometric integrators have been considered in the context of developing a numerical relativity based on the Regge calculus \cite{Frauendiener:2008bt,Richter:2008pr}, but the idea does not appear to have been fully pursued. We argue that such schemes merit full consideration for the reasons given above. 

In previous work \cite{markakis_time-symmetry_2019}, we applied a class of time-symmetric methods derived from Hermite integration to both the mechanics of a single particle and a classical wave equation sourced by a scalar charge. In the present work, we consider the master fields of black hole perturbation theory (BHPT), showing that for each field there are at least two conserved quantities derivable from Noether's theorem (energy and U(1) charge) and that Hermite methods numerically conserve both. We begin in Sec.~\ref{Hermite} by presenting an overview of method-of-lines numerics with Hermite integration, then proceed to the integration of classical fields in Sec.~\ref{PDEs}. We begin with the Schr\"{o}dinger field of nonrelativistic theory, which serves as a familiar example for outlining the machinery of more advanced problems. We proceed to the massless Klein-Gordon field governed by a scalar wave equation in both flat and Schwarzschild spacetimes. We finally examine gravitational perturbations to the Schwarzschild spacetime in the Newman-Penrose formalism governed by the Bardeen-Press-Teukolsky (BPT) or Regge-Wheeler-Zerilli (RWZ) equations. In each case, we examine which Noether-related constants are numerically conserved.

\section{Time-Symmetric Evolution with Hermite Integration}
\label{Hermite}
We consider the problem of numerically approximating solutions to partial differential equations (PDEs). Since the equations of BHPT are hyperbolic, we proceed using the \textit{Method of Lines}. That is, for a hyperbolic or parabolic PDE
\begin{equation}\label{eq:GenericHyperbolicPDE}
    \partial_t u(t,x) = \hat L (u(t,x))
\end{equation}
where $\hat L$ is a (possibly nonlinear) spatial differential operator, we proceed by approximating the field $u(t,x)$ on a discrete spatial grid $\mathbf{X}=\{x_i\}_{i=0}^N$ so that $u(t,x) \rightarrow \mathbf{u} (t) $. The components $u(t,x_i) := u_i(t)$ of the vector $\mathbf{u} (t)$ are the values of the field evaluated at the gridpoints. Heuristically, this converts the problem from a PDE in space-time variables $(t,x)$ to a  system of coupled ordinary differential equations (ODEs) in one time variable $t$,
\begin{equation}\label{eq:GenericODE}
      \frac{d \mathbf{u}}{d t}  =\mathbf{L}(\mathbf{u}).
\end{equation}
where  
the matrix operator $\mathbf{L}$ couples the set of ODEs. In this section, we will outline a method for evolving such systems via numerical integration
schemes symmetric under time-reversal.

\subsection{Hermite Integration}
Using the fundamental theorem of calculus, the differential equations~\eqref{eq:GenericODE} can be converted to a system of integral equations,
\begin{equation}\label{eq:GenericIntEq}
    \mathbf{u}(t_{n+1}) = \mathbf{u}(t_n) + \int_{t_n}^{t_{n+1}} \mathbf{f}(t) dt,
\end{equation}
with the integrand $\mathbf{f}(t) =\mathbf{L}(\mathbf{u}(t))$ treated as a function of time $t$. The problem has thus been reduced to evaluating the time integral in Eq.~\eqref{eq:GenericIntEq}. 

\subsubsection{1-point Taylor expansion}
Integrating a (1-point) Taylor expansion of $\mathbf{f}(t)$ about the initial time $t_n$ yields the approximant
\begin{equation} \label{eq:Taylorint}
 \int_{t_n}^{t_{n+1}} \mathbf{f}(t) dt =
\sum_{m=1}^l \frac{\Delta t^m }{m!} \mathbf{f}^{(m-1)}_n 
+\mathbf{R}_l
\end{equation}
with remainder
\begin{equation} \label{eq:TaylorintRem}
\mathbf{R}_l=\frac{{{\Delta
{t^{l+1}}{}}}}{{(l+1){{!}}}}\mathbf{f}^{(l)}(t), \quad t \in [t_n, t_{n+1}].
\end{equation}
Here, we denote the $m$-th derivative of $\mathbf{f}(t)$ at $t=t_n$ by
\begin{equation} \label{eq:fder}
    \mathbf{f}_n^{(m)} = \left. \frac{d^m \mathbf{f}(t)}{dt^m}\right\rvert_{t=t_n}.
\end{equation}
The time derivatives \eqref{eq:fder} may be determined exactly by recursively applying the  chain rule, $\mathbf{f}^{(m)}=\frac{\partial \mathbf{f}^{(m-1)}}{\partial \mathbf{u}}  \dot{\mathbf{ u}}$, with the last term substituted from the
equation of motion \eqref{eq:GenericODE}. This results in a single-step Taylor method.
Alternatively, the derivatives \eqref{eq:fder} may be treated as constant polynomial coefficients and eliminated by evaluating the Taylor approximant of $\mathbf{f}(t)$ at multiple points, resulting in a multi-step method, such as Runge-Kutta. These two approaches are equivalent for linear systems. In any case, it is evident from Eq.~\eqref{eq:Taylorint} that Runge-Kutta methods or 1-point Taylor expansions violate time-symmetry (that is, $\mathbb{Z}_2$ symmetry under time-reversal, $t_n \leftrightarrow t_{n+1}$, $dt \rightarrow -dt$) and fail to preserve the symplectic structure or Noether charges of  Hamiltonian systems. 
\subsubsection{2-point Taylor expansion}
A time-symmetric integration scheme can be obtained by approximating $\mathbf{f}(t)$ with a 2-point Taylor expansion or, equivalently, a 2-point \textit{Hermite interpolant:} an osculating polynomial constructed to match the values of $f$ and its derivatives at the endpoints $t_n$ and $t_{n+1}$. Integrating this osculating polynomial from $t_n$ to $t_{n+1}$ approximates the integral in Eq.~\eqref{eq:GenericIntEq}. This procedure is detailed in \cite{markakis_time-symmetry_2019}. For the present work, we quote the end result. Let us denote the $m$-th order time-derivative of $\mathbf{f}(t)$ at time $t_n$ by
\begin{equation}
    \mathbf{f}_n^{(m)} = \left. \frac{d^m \mathbf{f}(t)}{dt^m}\right\rvert_{t=t_n}.
\end{equation}
Integrating a Hermite interpolating polynomial which osculates derivatives up to order $l-1$ yields the \textit{generalized Hermite rule} \cite{lanczos_applied_1956,dyche_multiple_1956}:
\begin{equation}\label{eq:GeneralHermiteRule}
    \int_{t_n}^{t_{n+1}} \mathbf{f}(t) dt = \sum_{m = 1}^l c_{l m} ~\Delta t^{m} \Big( \mathbf{f}_n^{(m-1)} ~+~ (-1)^{m-1} \mathbf{f}_{n+1}^{(m-1)} \Big) + \mathbf{R}_l
\end{equation}
with the expansion coefficients given by
\begin{equation} \label{eq:clm}
    c_{l m} = \frac{l!(2l-m)!}{m!(2l)!(l-m)!}
\end{equation}
and the remainder given by
\begin{equation}  \label{eq:Rerror}
    \mathbf{R}_l
     = (-1)^l \frac{(l!)^2}{(2l+1)!(2l)!} \Delta t^{2l+1} \mathbf{f}^{(2l)}(t), \quad t \in [t_n, t_{n+1}].
\end{equation} 
Neglecting the remainder term, one can approximate the integral by summing $l$ terms on the right side of Eq.~\eqref{eq:GeneralHermiteRule}. The most important feature of this formula is its  $\mathbb{Z}_2$ symmetry under time-reversal ($t_n \leftrightarrow t_{n+1}$, $dt \rightarrow -dt$). In addition, the remainder term scales like $\Delta t^{2l+1}$: although the formula only contains terms up to $\Delta t^l$, it is accurate to $\mathcal{O}(\Delta t^{2l})$. Moreover, the numerical pre-factor in Eq.~\eqref{eq:Rerror} decreases much more rapidly with increasing $l$ compared to a 1-point Taylor expansion (cf. \cite{lanczos_applied_1956,dyche_multiple_1956}). That is, even if we compare methods of the same order, the truncation error in a 2-point Taylor expansion is several orders of magnitude lower than that of methods based on a 1-point Taylor expansion (such as the usual
Runge-Kutta methods).

In this work, we will mainly demonstrate conservation properties of second- and fourth-order time-symmetric integration rules, so we state them now. The choice $l=1$ yields the familiar trapezium rule,
\begin{equation}\label{eq:trapizium}
    \int_{t_n}^{t_{n+1}} \mathbf{f}(t) dt = \frac{\Delta t}{2}(\mathbf{f}_n + \mathbf{f}_{n+1}) + \mathcal{O}(\Delta t^3),
\end{equation}
which is accurate to second order. The choice $l=2$ yields the Hermite rule,
\begin{equation}\label{eq:Hermite}
    \int_{t_n}^{t_{n+1}} \mathbf{f}(t) dt = \frac{\Delta t}{2}(\mathbf{f}_n + \mathbf{f}_{n+1}) + \frac{\Delta t^2}{12}( \mathbf{\dot f}_n - \mathbf{\dot f}_{n+1}) +  \mathcal{O}(\Delta t^5),
\end{equation}
which is accurate to fourth order. Here, the overdot indicates a time derivative. The choice $l=3$ yields Lotkin's rule \cite{lotkinNewIntegratingProcedure1952}. Higher order generalizations can be obtained by substituting $l=4, \ 5, \
\dots$  into Eq.~\eqref{eq:GeneralHermiteRule} as detailed in \cite{markakis_time-symmetry_2019}.

If the schemes~\eqref{eq:GeneralHermiteRule} are applied to the integral equation \eqref{eq:GenericIntEq}, an implicit scheme is obtained to solve for $\mathbf{u}(t_{n+1})$. Moreover, since it is an implicit \textit{multi-derivative method} of the kind studied by Brown \cite{brown_multi-derivative_1973,brown_characteristics_1977}, it is \textit{unconditionally stable}. That is, there is no Courant limit on the timestep $\Delta t$. And, as a time-symmetric method, it has been shown to numerically conserve the energy and symplectic structure of Hamiltonian systems \cite{markakis_time-symmetry_2019}.

\subsection{Application to Systems of Partial Differential Equations}\label{sec:LinearEquations}

\subsubsection{Method of lines with time-symmetric discretization }
Although they possess desirable theoretical properties, implicit methods are generally require numerically solving nonlinear algebraic equations at every time step. If the original PDE system is linear, then it is possible to construct an explicit evolution scheme from these methods. We discuss how to do so now.

If the time integral 
\eqref{eq:GenericIntEq}
is approximated by the \textit{trapezium rule} \eqref{eq:trapizium}, we have
\begin{equation}\label{eq:LD2}
    \mathbf{u}^{n+1} = \mathbf{u}^n + \frac{\Delta t}{2}[ \mathbf{L} ( \mathbf{u}^n) + \mathbf{L} ( \mathbf{u}^{n+1} )].
\end{equation}
This amounts to the Crank-Nicolson scheme, generalized here for any spatial discretization.  Eq.~\eqref{eq:LD2} is implicit and Choptuik has suggested that it may be solved by self-consistent iteration (that is, using an initial guess $ \mathbf{u}^{n+1}= \mathbf{u}^{n}$ on the right side, computing a new value $\mathbf{u}^{n+1}$ on the left using the above equality, substituting that new value on the right side, and so forth, iterating until convergence is achieved). This approach is straightforward and also applicable to non-linear systems, that is, even when $\hat L$ is a non-linear operator. (One might also use Newton-Raphson iteration, albeit, for sufficiently small time-steps, self-consistent iteration is rapidly convergent and simpler to implement.) It has been argued that this scheme should be iterated twice and not more \cite{Teukolsky:1999rm}, as iterating more than twice does not improve the stability or the formal order of the scheme. However, the nuance of time symmetry has been lost in this argument: iterating only twice violates the time-symmetry
inherent in  Eq.~\eqref{eq:LD2}, leading to numerical violation of energy and symplectic structure \cite{Richter:2008pr}.  
In a companion paper, it will be shown that more iterations  effectively restore time-symmetry and conserve Noether charges and symplectic structure, leading to better behavior in long-time numerical simulations.
Here, we will focus on linear PDEs, whence Eq.~\eqref{eq:LD2} and its higher order generalizations can be solved via direct matrix inversion, which effectively amounts to "infinite" iterations, and preserves time-symmetry, symplecticity and most (but not all) Noether charges to machine epsilon.

If we specify that the differential operator $\hat L$ in Eq. \eqref{eq:GenericHyperbolicPDE} is in fact linear, then its spatial discretization is also linear:
\begin{equation}  \label{eq:Ldotu}
   \hat L u(t,x) \rightarrow (\hat L u(t,x) )_\imath = \sum_\jmath {L}_{\imath \jmath} u_\jmath(t).
\end{equation}
This is to say that the differential operator $\hat L$, upon discretization, amounts to a matrix $\mathbf{L}$ which then acts on a vector $\mathbf{u}(t)$ representing the discrete approximation to the field $u(t,x)$. The differential
equation \eqref{eq:GenericODE} becomes
\begin{equation} \label{eq:GenericLinearODE}
    \frac{d \mathbf{u}}{d t} = \mathbf{L} \cdot \mathbf{u}
\end{equation}
where a dot $\cdot$ denotes summation over adjacent indices, that is, matrix-vector multiplication in the case of Eq.~\eqref{eq:Ldotu}. Then, the integral equation \eqref{eq:GenericIntEq} becomes
\begin{equation}   \label{eq:GenericIntEqLu}
    \mathbf{u}^{n+1} = \mathbf{u}^n + \mathbf{L} \cdot \int_{t_n}^{t_{n+1}}
\mathbf{u}(t)~ dt
\end{equation}
where $\mathbf{u}^{n}:=\mathbf{u}({t_n})$ and we assumed that $\hat L$ is time-independent. Various collocation methods for computing the discretized spatial differentiation operator  $\mathbf{L}$ will be discussed below. In this section, we discuss symmetric discretizations in time.

If the operator $\hat L$ is linear, iteration may be avoided by using matrix inversion. If using finite differencing for spatial discretization, then the matrix $\mathbf{L}$ is sparse, and one can write the above scheme as
\begin{equation}\label{eq:LD2OperatorThomas}
    \bigg( \mathbf{I} - \frac{\Delta t}{2} \mathbf{L} \bigg) \cdot \mathbf{u}^{n+1} = \bigg( \mathbf{I} + \frac{\Delta t}{2} \mathbf{L} \bigg) \cdot \mathbf{u}^n,
\end{equation}
where $\mathbf{I}$ is the identity matrix. One can then solve for $\mathbf{u}^{n+1} $ using the tridiagonal matrix (Thomas) algorithm, or its variants, in $\mathcal{O}(N)$ operations. Parallel tridiagonal solvers have been developed for many vector and parallel architectures, including GPUs, making this option efficient.

If using pseudo-spectral methods (such as Fourier or Chebyshev collocation methods) for spatial discretization, then the matrix $\mathbf{L}$ is full. Direct  matrix inversion may be used to solve for $\mathbf{u}^{n+1}$ explicitly,
\begin{equation}\label{eq:LD2Operator}
    \mathbf{u}^{n+1} = \bigg( \mathbf{I} - \frac{\Delta t}{2} \mathbf{L} \bigg)^{-1} \cdot \bigg( \mathbf{I} + \frac{\Delta t}{2} \mathbf{L} \bigg) \cdot \mathbf{u}^n,
\end{equation}
at the cost of $\mathcal{O}(N^3)$ operations. Although the matrices involved are full, for linear PDEs they are constant throughout the evolution. Thus, the matrix inverse may be stored in memory,   reducing the numerical evolution to a simple matrix multiplication, costing $\mathcal{O}(N^2)$ operations per time step. The condition number of the matrices to be inverted is typically low, so  inversion does not entail significant loss of precision. To reduce round-off error in each time step, the above scheme can be written in the more numerically precise form:
\begin{equation}\label{eq:LD2OperatorPrecise}
    \mathbf{u}^{n+1} = \mathbf{u}^n+\bigg( \mathbf{I} - \frac{\Delta t}{2} \mathbf{L} \bigg)^{-1} \cdot (\Delta t \, \mathbf{L})  \cdot \mathbf{u}^n.
\end{equation}
Although we are not using compensated summation,  the scheme \eqref{eq:LD2OperatorPrecise} often conserves Noether charges to machine epsilon, while the (analytically equivalent) scheme \eqref{eq:LD2Operator} accumulates round-off error over time.

A fourth-order scheme follows from the \textit{Hermite rule} \eqref{eq:Hermite}:
\begin{equation}\label{eq:LD4}
    \mathbf{u}^{n+1} = \mathbf{u}^n + \frac{\Delta t}{2} \mathbf{L} \cdot \big( \mathbf{u}^n +  \mathbf{u}^{n+1} \big) +  \frac{\Delta t^2}{12} \mathbf{L} \cdot \big( \Dot{\mathbf{u}}^n -  \Dot{\mathbf{u}}^{n+1} \big) .
\end{equation}
One can invoke the differential equation \eqref{eq:GenericLinearODE} to replace time derivatives $\Dot{\mathbf{u}}$ by spatial derivatives $ \mathbf{L} \cdot \mathbf{u}$ and then solve for $\mathbf{u}^{n+1}$ using the methods outlined above.  If the matrix $\mathbf{L}$ is full, this may be done explicitly using matrix inversion:
\begin{equation}\label{eq:LD4Operator}
    \mathbf{u}^{n+1} = \bigg( \mathbf{I} - \frac{\Delta t}{2} \mathbf{L} + \frac{\Delta t^2}{12} \mathbf{L}^2 \bigg)^{-1} \cdot \bigg( \mathbf{I} + \frac{\Delta t}{2}\mathbf{L} + \frac{\Delta t^2}{12} \mathbf{L}^2 \bigg) \cdot \mathbf{u}^n
\end{equation}
or, equivalently,
\begin{equation}\label{eq:LD4OperatorPrecise}
    \mathbf{u}^{n+1} = \mathbf{u}^n + \bigg[ \mathbf{I} - \frac{\Delta t}{2} \mathbf{L} \cdot \bigg(\mathbf{I}-\frac{\Delta t}{6} \mathbf{L}  \bigg) \bigg]^{-1} \cdot (\Delta t \, \mathbf{L})  \cdot \mathbf{u}^n,
\end{equation}
with the scheme \eqref{eq:LD4OperatorPrecise} being superior in terms of reducing round-off error. The Hermite scheme~\eqref{eq:LD4OperatorPrecise} is used to produce all numerical results in this paper.

In the cases discussed above, as well as higher order cases, evolving the system by one time step amounts to matrix-vector multiplication and addition (MMA) of the general form
\begin{equation} \label{eq:LDlSchemePrecise}
    \mathbf{u}^{n+1} = \mathbf{A} \cdot \mathbf{u}^n = \mathbf{u}^n+(\mathbf{A}-\mathbf{I}) \cdot \mathbf{u}^n,
\end{equation}
with the trapezium evolution matrix
\begin{equation} \label{eq:A2}
    \mathbf{A}^{(2)} = \bigg( \mathbf{I} - \frac{\Delta t}{2} \mathbf{L} \bigg)^{-1} \cdot \bigg( \mathbf{I} + \frac{\Delta t}{2} \mathbf{L}
    \bigg),
\end{equation}
and the Hermite evolution matrix
\begin{equation} \label{eq:A4}
    \mathbf{A}^{(4)} = \bigg( \mathbf{I} - \frac{\Delta t}{2} \mathbf{L} + \frac{\Delta t^2}{12} \mathbf{L}^2 \bigg)^{-1} \cdot \bigg( \mathbf{I} + \frac{\Delta t}{2} \mathbf{L} + \frac{\Delta t^2}{12} \mathbf{L}^2 \bigg),
\end{equation}
corresponding to Eqs.~\eqref{eq:LD2Operator}-\eqref{eq:LD2OperatorPrecise} and \eqref{eq:LD4Operator}-\eqref{eq:LD4OperatorPrecise} respectively.
The last equality of the evolution scheme~\eqref{eq:LDlSchemePrecise}, which separates out the change in $\mathbf{u}$ in each time step, entails \textit{significantly lower round-off error}, and constitutes our scheme of choice. Because MMA operations are parallelized by modern CPU and GPU libraries, this scheme requires  little programming to implement efficiently.

Extending the scheme to higher order in time is straightforward. Using the generalized Hermite rule \eqref{eq:GeneralHermiteRule}, and using
the equation of motion \eqref{eq:GenericLinearODE} repeatedly to replace time derivatives $d^m \mathbf{u} /dt^m$ with spatial derivatives $\mathbf{L}^m \cdot \mathbf{u}$, one can obtain a $2l^{\rm th}$-order approximation to the \textit{evolution matrix}:
\begin{equation}   \label{eq:LDlA}
    \mathbf{A}= \bigg[ \sum_{m = 0}^l c_{l m} ~(-\Delta t \, \mathbf{L})^{m}\bigg]^{-1} \cdot \bigg[ {\sum_{m = 0}^l c_{l m} ~(\Delta t \, \mathbf{L})^{m}}\bigg]
\end{equation}   
with the coefficients $c_{lm}$ given by Eq.~\eqref{eq:clm}. The class of schemes \eqref{eq:LDlSchemePrecise}-\eqref{eq:LDlA} is  accurate to $\mathcal{O}(\Delta t^{2l})$ and manifestly time-symmetric, that is, invariant under the exchange $t_{n+1}\leftrightarrow t_n$. It will be demonstrated that this vital property leads to \textit{numerical conservation of certain Noether charges} of the system \eqref{eq:GenericHyperbolicPDE}. Moreover, by construction, the spectral radius of the evolution matrix is
$\rho(\mathbf{A})=1$ for any spatial discretization that satisfies appropriate boundary conditions. Therefore this class of time-symmetric schemes is \textit{unconditionally stable}.

As mentioned earlier, when the matrix $\mathbf{L}$ is full, fast sparse array algorithms are inapplicable. Numerical evolution via the scheme~\eqref{eq:LDlSchemePrecise} is then more precise and efficient by performing a matrix inversion, computing the matrix
\begin{equation}  \label{eq:LDlAminusI}
    \mathbf{A}-\mathbf{I} = \bigg[ \sum_{m = 0}^l c_{l m} ~(-\Delta t \, \mathbf{L})^{m}\bigg]^{-1} \cdot \bigg[ {\sum_{\substack{m = 1 \\ m \text{ odd}}}^l 2 c_{l m} ~(\Delta t \, \mathbf{L})^{m}}\bigg]
\end{equation}
in advance, storing it in memory, and using it to perform a MMA in each time step, as dictated by the last equality of Eq.~\eqref{eq:LDlSchemePrecise}. Expressing all matrix polynomials in Horner form with respect to $\Delta t \, \mathbf{L}$ reduces round-off error in computing the matrix $\mathbf{A}-\mathbf{I} $. Since we use Fourier and Chebyshev collocation methods for spatial discretization in this paper, this is the scheme we opt for.  Substituting $l=1$ or $l=2$ to the above equation recovers the trapezium and Hermite rule schemes discussed earlier. In general, due to its unconditional stability, Noether-charge preserving properties, and very low truncation and roundoff errors, the time-symmetric scheme~\eqref{eq:LDlSchemePrecise}-\eqref{eq:LDlAminusI} is well suited for long time numerical evolution in black hole perturbation theory.

\subsubsection{Relation to Pad\'e Approximants}
In the case of linear systems, the evolution schemes derived in this section may also be derived by the method of Pad\'e expansions. Note that Eq.~\eqref{eq:GenericLinearODE} has a formal solution using matrix exponentiation, $\mathbf{u}(t) = e^{t\, \mathbf{L}} \cdot \mathbf{u}(0)$, which can be used to integrate Eq.~\eqref{eq:GenericIntEqLu} exactly in time:
\begin{equation}
    \mathbf{u}^{n+1} = e^{\Delta t \, \mathbf{L}} \cdot \mathbf{u}^n.
\end{equation}
Approximate schemes may then be obtained by expanding the exponential in
powers of $\Delta t$. A one-point Taylor expansion,
\begin{equation} \nonumber
    e^{\Delta t \, \mathbf{L}} \simeq \sum_{m=0}^l \frac{1}{m!}(\Delta t \, \mathbf{L})^m,
\end{equation}
is equivalent to a classical Runge-Kutta scheme of order $l$. If instead one uses a symmetric Pad\'e expansion, approximating the exponential with a rational function of polynomials, 
\begin{equation} \nonumber
    e^{\Delta t \, \mathbf{L}} \simeq \bigg[ \sum_{m = 0}^l c_{l m} ~(-\Delta t \, \mathbf{L})^{m}\bigg]^{-1} \cdot {\sum_{m = 0}^l c_{l m} ~(\Delta t \, \mathbf{L})^{m}},
\end{equation}
one recovers the time-symmetric formula~\eqref{eq:LDlA} obtained earlier via Hermite integration.

Although both methods are equivalent in the linear case, Hermite integration methods tend to generalize better. First, a formal solution in terms of an operator exponential, which  Pad\'e methods rely upon, is only valid for linear operators $\hat L.$ The generalization to nonlinear operators is unclear (albeit certain non-linearities can be  accommodated via Duhamel's principle \cite{2018JCoPh.356}).
By contrast, Hermite integration does not \textit{a priori} assume linearity: the resulting systems of equations can be easily solved via self-consistent (or Newton-Raphson) iteration for non-linear systems.  Second, Hermite integration can  accommodate distributional source terms added to Eq.~\eqref{eq:GenericHyperbolicPDE}. In particular, discontinuous Hermite rules can be obtained with the method of undetermined coefficients, which can be  generalized  to accommodate jump discontinuities across distributional sources (see \cite{Markakis:2014nja} for details). Distributional source terms  arise, for instance, in the  motion of a particle orbiting a black hole. Thus, the class of time-symmetric Hermite integration schemes used in this paper can be  applied to EMRIs and used to compute the gravitational self-force in the time domain.

\subsubsection{Spatial discretization with collocation methods}

For the PDE systems considered here, the differential operator $\hat L$ is generally a combination of first and/or second order partial derivatives with respect to  spatial coordinates. Thus, in a method of lines context, the matrix $\mathbf{L}$ will be a combination of differentiation matrices $\mathbf{D}^{(1)}$ and/or  $\mathbf{D}^{(2)}$ respectively. That is, upon discretization with collocation methods, spatial differentiation will amount to matrix-vector multiplication:
\begin{eqnarray*}
    \partial_x u \big|_{x=x_\imath} \simeq ~ \sum_\jmath {D}^{(1)}_{\imath \jmath} u_\jmath, \quad \partial^2_x u \big|_{x=x_\imath} \simeq ~ \sum_j {D}^{(2)}_{\imath \jmath} u_\jmath
\end{eqnarray*}
\textit{Finite differences}\\

A $2^\text{nd}$-order finite-difference method gives the differentiation matrix
\begin{eqnarray}
    {D}^{(1)}_{\imath \jmath}
    &=&\frac{{{x_\imath} - {x_{\imath - 1}}}}{{({x_{\imath + 1}} - {x_\imath})({x_{\imath + 1}} - {x_{\imath- 1}})}}{\delta _{\jmath,\imath + 1}} + \frac{{{x_{\imath + 1}} + {x_{\imath - 1}} - 2{x_i}}}{{({x_{\imath + 1}} - {x_\imath}) ({x_\imath} - {x_{\imath - 1}})}}{\delta _{\imath \jmath}}  \\
    &-&  \frac{{{x_{\imath + 1}} - {x_\imath}}}{{({x_{\imath + 1}} - {x_{\imath - 1}})({x_\imath} - {x_{\imath - 1}})}}{\delta _{\jmath,\imath - 1}} \nonumber  
\end{eqnarray}
for all $\imath=1,\dots,N-1$, while one-sided finite differences are used  at the end-points $\imath=0,N$. The finite-difference error is $\mathcal{O}(\Delta x ^ { 2 } )$, where $\Delta  x $ is the maximum local grid spacing. Fast methods for  computing the  matrices have been developed \cite{Fornberg1998,Fornberg1988,Fornberg2006,Sadiq2011,Welfert1997} and implemented in computational libraries. The Wolfram Language command
\begin{verbatim}
    D1=NDSolve`FiniteDifferenceDerivative[Derivative[1],X,
    "DifferenceOrder"->2]@"DifferentiationMatrix"
\end{verbatim}
uses Fornberg's algorithm \cite{Fornberg1998,Fornberg1988,Fornberg2006} to compute the  $1$-st order differentiation matrix $\mathbf{D}^{(1)}$ with  $2$-nd order finite differencing for a given list of nodes $\texttt{X} =\{x_0,x_1,x_2,...,x_N\}$. Spatial integrals may be computed using, for instance, the trapezium rule \cite{wolfram}.

\hfill \break
\textit{Chebyshev collocation methods} \newline

Exponential convergence may be attained using a Chebyshev pseudo-spectral method, where spatial grid points are the extrema of the $N^{\text{th}}$ order Chebyshev polynomial. In the interval $x\in[a,b]$, the Chebyschev-Gauss-Lobatto nodes are given by
\begin{equation}  \label{ChebyshevNodes}
    x_\imath=\frac{{{b+a} }}{2} + \frac{{{b} - {a}}}{2} z_\imath, \quad z_\imath = \sin \bigg(\frac{2\imath-N}{2N} \pi\bigg),\quad \imath=0,1,\dots,N
\end{equation}
The first derivative operator on this grid is
\begin{equation}
    {D}^{(1)}_{\imath \jmath} = 
\frac{2}{b-a}    \begin{cases}
        \frac{c_\imath (-1)^{\imath+\jmath}}{c_\jmath (z_\imath-z_\jmath)} & \imath \neq \jmath\\
        -\frac{z_\jmath}{2(1-z_\jmath^2)} & \imath = \jmath \neq 0,N\\
        -\frac{2 N^2 + 1}{6} & \imath = \jmath = 0\\
        \frac{2 N^2 + 1}{6} & \imath = \jmath = N
    \end{cases}
\end{equation}
where $c_0 = c_N = 2$ and $c_1, \dotsc, c_{N-1} = 1$. 

The second derivative
operator can be evaluated by $ \mathbf{D}^{(2)} = (\mathbf{D}^{(1)} )^2$
or, equivalently \cite{canuto_2006},
\begin{equation}\label{ChebyshevD2}
D_{\imath \jmath}^{(2)} = {\left( {\frac{2}{{b - a}}} \right)^2}\left\{ {\begin{array}{*{20}{l}}
{\frac{{{{( - 1)}^{\imath + \jmath}}}}{{{c_\jmath}}}\frac{{z_\imath^2 + {z_\imath}{z_\jmath} - 2}}{{(1 - z_\imath ^2){{({z_\imath} - {z_j})}^2}}}}&{\imath \ne \jmath,{\;\;\; }\imath \ne 0,\,{\;\;\; }\imath \ne N}\\
{\frac{2}{3}\frac{{{{( - 1)}^\jmath}}}{{{c_j}}}\frac{{(2{N^2} + 1)(1 + {z_j}) - 6}}{{{{(1 + {z_j})}^2}}}}&{i \ne j,{\;\;\; }\imath = 0}\\
{\frac{2}{3}\frac{{{{( - 1)}^{\jmath + N}}}}{{{c_\jmath}}}\frac{{(2{N^2} + 1)(1 - {z_\jmath}) - 6}}{{{{(1 - {z_\jmath})}^2}}}}&{\imath \ne \jmath,{\;\;\; }\imath = N}\\
{ - \frac{{({N^2} - 1)(1 - z_\jmath^2) + 3}}{{3{{(1 - z_\jmath^2)}^2}}}}&{\imath = \jmath,\,{\;\;\; }\imath \ne 0,{\;\;\; }\imath \ne N}\\
{\frac{{{N^4} - 1}}{{15}}}&{\imath = \jmath = 0{\;\;\; \rm{ or }\;\;\; }N}
\end{array}} \right.
\end{equation}
The Chebyshev differentiation matrices can be constructed automatically via
the Wolfram Language commands:
 \begin{verbatim}
D1=NDSolve`FiniteDifferenceDerivative[Derivative[1],X,"DifferenceOrder"->
"Pseudospectral"]@"DifferentiationMatrix"
\end{verbatim}
 \begin{verbatim}
D2=NDSolve`FiniteDifferenceDerivative[Derivative[2],X,"DifferenceOrder"->
"Pseudospectral"]@"DifferentiationMatrix"
\end{verbatim}
which respectively return the  matrices $\mathbf{D}^{(1)},\mathbf{D}^{(2)}$ for a list $\texttt{X}$ of nodes given by Eq.~\eqref{ChebyshevNodes}. Spatial integrals on this grid may be computed via Clenshaw-Curtis quadrature
\begin{equation}
    \int_a^b {f(x)dx}  \simeq  \sum\limits_{\imath=0}^{N} w_{\imath}f(x_{\imath}),{\rm{}} \quad w_\imath=\frac{b-a}{2}
    \begin{cases}
        \frac{1}{N^{2}}, & \imath=0,N  \\
        \frac{2}{N}\left(1-\sum_{k=1}^{(N-1)/2}\frac{2 \cos(2k \theta_\imath )}{4k^2-1} \right) & 0<\imath<N \\
    \end{cases}{}
    \quad  
\end{equation}
on a grid of an even number of points ($N$ odd), where 
$\theta_\imath=\imath \pi/N$.

\hfill \break
\textit{Fourier collocation methods} 
\newline

Chebyshev spectral differentiation matrices are not normal. For problems allowing periodic boundary conditions, we will use Fourier spectral differentiation matrices which are symmetric or skew-symmetric. For equidistant nodes in the interval $x \in  [a,b)$, 
\begin{equation} \label{EquidistantNodes}
    x_\imath=a+\imath\frac{{b- {a}}}{N}, \quad \imath=0,1,\dots,N,
\end{equation}
the first derivative matrix is given by
\begin{equation} \label{FourierD1}
    {D}^{(1)}_{\jmath k} = \frac{2 \pi}{b - a} 
    \begin{cases}
        0 & \jmath = k\\
        \frac{1}{2} (-1)^{\jmath + k} \cot \Big( \frac{\pi (\jmath - k) }{N} \Big) & \jmath \neq k\\
    \end{cases}
\end{equation}
This matrix is skew-symmetric by construction.
The second derivative matrix in this scheme takes different forms depending on whether $N$ is odd or even. We will take $N$ to be even in this work, whence
\begin{equation} \label{FourierD2}
    {D}^{(2)}_{\jmath k} = \bigg( \frac{2 \pi}{b - a} \bigg)^2
    \begin{cases}
        - \frac{N^2+2}{12}  & \jmath = k\\
        - \frac{1}{2}(-1)^{\jmath - k}  \csc^2 \Big( \frac{\pi (\jmath - k) }{N} \Big) & \jmath \neq k\\
    \end{cases}.
\end{equation}
This matrix is symmetric by construction. The Fourier differentiation matrices can be constructed automatically via the Wolfram Language commands:
\begin{verbatim}
    D1=NDSolve`FiniteDifferenceDerivative[Derivative[1],X,"DifferenceOrder"->
    "Pseudospectral",PeriodicInterpolation->True]@"DifferentiationMatrix"
\end{verbatim}
\begin{verbatim}
    D2=NDSolve`FiniteDifferenceDerivative[Derivative[2],X,"DifferenceOrder"->
    "Pseudospectral",PeriodicInterpolation->True]@"DifferentiationMatrix"
\end{verbatim}
which respectively return the  matrices $\mathbf{D}^{(1)},\mathbf{D}^{(2)}$ for a list $\texttt{X}$ of equidistant nodes \eqref{EquidistantNodes}. For this periodic grid, spatial integrals can be computed via the trapezium rule, which is exponentially convergent for periodic functions.

\section{Numerical conservation of Noether charges}
\label{PDEs}
In this section, we apply Hermite integration to the study of complex-valued scalar (and scalar-like) fields. Such quantities are of immense utility in classical field theories and have important applications in relativistic astrophysics. When such theories are posed as variational principles, it becomes a straightforward matter to determine quantities which are conserved in evolution. They can be obtained either from Noether's theorem, relating continuous symmetries of the action functional correspond to conserved quantities (``Noether charges''), or introducing canonical variables and posing the problem in Hamiltonian dynamics, revealing an underlying geometric structure (symplecticity) and an associated differential volume form that is preserved. The degree to which a numerical method conserves these quantities in evolution presents an important test of the method's strength: if such quantities are not conserved, then the physical validity of a simulation's results is called into question.

We argue that Hermite integration schemes are strong candidates for respecting these conservation laws and provide numerical evidence demonstrating that they do in simulation, provided that appropriate boundary conditions are implemented, and spatial differentiation error and round-off error are minimized. We first explore the Schr\"odinger wavefunction of nonrelativistic quantum mechanics and the massless Klein-Gordon field of classical field theory as prototypical examples of the properties, and then we discuss the Regge-Wheeler-Zerilli and Bardeen-Press-Teukolsky fields arising in BHPT.

\subsection{The Schr\"odinger Field}
The equation for the Schr\"odinger field $\psi$ of a quantum particle in a one-dimensional potential $V(t,x)$ reads
\begin{equation}\label{eq:SchrodingerOriginal}
   {\rm{i}\,} \hbar \partial_t \psi = - \frac{\hbar^2}{2 m} \partial_x^2 \psi + V \psi,
\end{equation}
where ${\rm{i}\,}$ denotes the imaginary unit. This classical field equation is of the first-order in time form~\eqref{eq:GenericHyperbolicPDE}, and can be written as 
\begin{equation}\label{eq:1stELsimple}
    \partial_t  \psi = \hat L  \psi.
\end{equation}
where $\hat L = \frac{{\rm{i}\,}\hbar}{2 m} \partial_x^2 -{\rm{i}\,} V$ is a linear operator.
Thus, the numerical schemes outlined above for Eq.~\eqref{eq:GenericHyperbolicPDE} are readily applicable. 

To compare the conservation properties of numerical schemes, we begin by deriving Noether charges related to gauge and Galilean symmetries of the Schr\"odinger equation.
It is possible to write down a classical action for which Eq.~\eqref{eq:SchrodingerOriginal} is the Euler-Lagrange equation. There are several choices (Lagrangian and action functionals are nonunique), but we find an action first-order in space and time,
\begin{equation} \label{eq:1stOrderAction}
    S[\psi,\psi^\star] =\int dt dx \ \mathcal{L} =\int dt dx \ \bigg[  \frac{{\rm{i}\,} \hbar}{2} ( \psi^\star \partial_t \psi - \psi \partial_t \psi^\star) - \frac{\hbar^2}{2 m} \partial_x \psi ~ \partial_x \psi^\star - V \psi^\star \psi \bigg],
\end{equation}
to be the most convenient for deriving conserved quantities.
Notice that this action is real and treats the classical field $\psi$ and and its complex conjugate $\psi^\star$ equally as independent quantities. Extremizing the action functional~\eqref{eq:1stOrderAction} with respect the complex (or Hermitian) conjugate $\psi^\star$ yields the field equation~\eqref{eq:SchrodingerOriginal}
\footnote{Varying with respect to $\psi$ would yield the complex conjugate of Eq. \eqref{eq:SchrodingerOriginal}}. 
Using the Euler-Lagrange equations,
\begin{equation}\label{eq:SchrodingerActionEL}
    \frac{\partial \mathcal{L} }{\partial  \psi^\star} - \partial_\mu p^\mu =0, \quad p^\mu = \frac{\partial \mathcal{L}}{\partial (\partial_\mu \psi^\star)},
\end{equation}
where the repeated index $\mu$ implies summation over the $t$ and $x$ coordinates, it is a simple matter to confirm that Eq.~\eqref{eq:SchrodingerOriginal} follows from extremizing the action~\eqref{eq:1stOrderAction}. The advantage of such a formulation is that it is now possible to form conserved quantities via Noether's theorem. Heuristically, Noether's theorem states that there is a conserved current for each continuous symmetry of the action.
We consider two kinds of symmetries here: those of the field (internal symmetries) and those of the potential (external symmetries).

If a continuous infinitesimal transformation $\psi \rightarrow \psi + \delta \psi$ leaves the action Eq.~\eqref{eq:1stOrderAction} unchanged or, equivalently, changes the Lagrangian density by a total divergence $\delta \mathcal{L} = \partial_\mu \delta A^\mu $, then the Noether current \cite{toth_noether_2018} (in spacetime notation)
\begin{equation}\label{eq:1stNoetherCurrent}
    J^\mu = p^\mu \delta \psi^\star  +  p^{\star \mu} \delta \psi - \delta A^\mu
\end{equation}
is conserved:
\begin{equation}\label{eq:1stNoetherTheorem}
    \partial_\mu J^\mu = \delta \mathcal{L} - \partial_\mu \delta A^\mu=0.
\end{equation}
Integrating this local conservation law over a spacelike surface $\Sigma_t$, which extends to spatial (or null) infinity, invoking Gauss' theorem and dropping boundary terms (valid when the field is supported only in the interior of $\Sigma_t$), yields a global conservation law: 
\begin{equation}
    \frac{d}{dt}\int_{\Sigma_t} dx \, J^t =0
\end{equation}
for the total Noether charge. 

The Schr\"{o}dinger field possesses an internal symmetry, stemming from invariance action~\eqref{eq:1stOrderAction} with respect to the $\delta\alpha$-family of global U(1) complex phase rotations, $\psi \rightarrow e^{{\rm{i} \,} \,  \delta\alpha} \psi$ where $ \delta \alpha \in \mathbb{R}$ is a constant parameter. This implies the infinitesimal transformation $\psi \rightarrow \psi +{\rm{i} \,} \delta \alpha \psi$ leaves Eq. \eqref{eq:1stOrderAction} unchanged (to linear order in $\delta \alpha$). Thus, the current
\begin{equation}
    J^\mu = ( \psi^\star \psi ,~ \psi^\star \partial_x \psi - \psi \partial_x \psi^\star )
\end{equation}
is locally conserved and the U(1) charge
\begin{equation}\label{eq:1stCharge}
    Q  = \int_{\Sigma_t} dx \ \psi^\star \psi.
\end{equation}
is globally conserved.

In addition, the Schr\"{o}dinger equation may have external symmetries, depending on the symmetries of the potential $V(t,x)$.
For a constant potential, the Schr\"{o}dinger equation \eqref{eq:SchrodingerOriginal} is symmetric under the Galilei group Gal(3), representing space-time translations or boosts (defined via the Galilean notion of space-time).
For the action~\eqref{eq:1stOrderAction}, the energy of the Schr\"{o}dinger field:
\begin{equation}
    E = 
    \int_{\Sigma_t} dx \bigg( \frac{\hbar^2}{2 m} \partial_x \psi ~ \partial_x \psi^\star + V \psi^\star \psi \bigg).
\end{equation}
is Noether-related to symmetry under time translations,
that is, the energy is conserved if the potential is time-independent, $\partial_t V = 0$.

The linear momentum of the Schr\"{o}dinger field:
\begin{equation}
    P = 
    \frac{{\rm{i}\,} \hbar}{2} \int_{\Sigma_t} dx ( \psi^\star \partial_x \psi - \psi \partial_x \psi^\star ).
\end{equation}
is Noether-related to symmetry under spatial translations,
that is, linear momentum is conserved if the potential is spatially constant potential, $\partial_x V = 0$.

If the potential $V$ is (spatially and temporally) constant, then a Galilean boost $x \rightarrow x - v t$ (this is a velocity boost appropriate for the present notion of spacetime, for some constant $v$) also leaves the equation of motion unchanged. It follows that the ``center of mass'' charge of the system,
\begin{equation}
    C = \int_{\Sigma_t} dx ( -{\rm{i}\,} t \hbar \psi^\star \partial_x \psi - m \psi^\star x \psi )
\end{equation}
is also a constant of motion\footnote{Conservation of this charge indicates that the center of mass moves with constant velocity. Observe that this charge, defined in terms of the classical field $\psi$, amounts to the expectation value $\langle p t - m x \rangle = m\langle v t - x \rangle$ in nonrelativistic quantum theory.}.

Like other classical field theories, the Schr\"{o}dinger field is endowed with symplectic structure. To reveal it, we first define a canonical momentum for the field:
\begin{equation}\label{eq:1stCanonicalMomenta}
    \pi := p^t = \frac{\hbar}{2 {\rm{i}\,}} \psi, \quad \pi^\star  := p^{\star t} = \frac{{\rm{i}\,} \hbar}{2} \psi^\star
\end{equation}
We construct a Hamiltonian density by the Legendre transformation
\begin{equation}\label{eq:SchroHamiltonian}
    \mathcal{H} = \pi^\star \partial_t \psi + \pi \partial_t \psi^\star - \mathcal{L} = \frac{2}{m} \partial_x \pi ~ \partial_x \pi^\star + \frac{4}{\hbar^2} V \pi^\star \pi
\end{equation}
This Hamiltonian density is unusual since it contains explicit dependence on $\partial_x \pi$. (This peculiarity stems from the fact that the action contains no terms quadratic in $\partial_t \psi$, so the canonical momentum is just the field itself.) It is thus necessary to modify the usual Hamilton equations of motion, and this is done in Appendix \ref{sec:NewHamilton}. When applied to this Hamiltonian, the canonical equations read
\begin{equation}
    \partial_t \psi = \frac{\partial \mathcal{H}}{\partial \pi^\star} - \partial_x \bigg( \frac{\partial \mathcal{H}}{\partial ( \partial_x \pi^\star)} \bigg) = \frac{4}{\hbar^2} V \pi - \frac{2}{m} \partial_x^2 \pi
\end{equation}
\begin{equation}
    \partial_t \pi = - \frac{\partial \mathcal{H}}{\partial \psi^\star} + \partial_x \bigg( \frac{\partial \mathcal{H}}{\partial (\partial_x \psi^\star)} \bigg) = \frac{\hbar^2}{2 m} \partial_x^2 \psi - V \psi 
\end{equation}
With a Hamiltonian formulation, the field $\psi$ and canonical momentum $\pi$ define a symplectic geometry.

There is one more consideration for numerical evolution of the Schr\"{o}dinger equation. If $\hat L$ is independent of time, then Eq. \eqref{eq:1stELsimple} possesses the formal solution $\psi(t,x) = e^{t \hat L} \psi(0,x)$. In nonrelativistic quantum theory, $\hat L$ is a Hermitian operator ($\hat L^\dagger = \hat L$). So, the exponential operator in this solution is \text{unitary}. (A direct consequence of unitarity is that the U(1) charge \eqref{eq:1stCharge} is  conserved). Ideally, a numerical approximation to $\psi(t,x)$ should also preserve the U(1)  charge (or its discrete analog).

For suitable spatial discretizations, the Hermite methods (unlike Runge-Kutta methods) described in the previous section are unitary. For this problem, the trapezium 
evolution matrix \eqref{eq:A2}, 
the Hermite  evolution matrix \eqref{eq:A4}
and higher order generalizations given by Eq.~\eqref{eq:LDlA} are, by construction, unitary:
\begin{equation}
     \mathbf{A}^\dagger \cdot \mathbf{A} = \mathbf{I}
\end{equation}
provided that the matrix $\mathbf{L}$ is Hermitian. As discussed earlier, upon discretization via collocation methods, this differential operator will amount to a matrix. Then,  Hermiticity requires a spatial discretization such that the second derivative matrix is symmetric. Examples of such schemes include  finite-difference methods, the Fourier pseudo-spectral method, and the Whittaker-Shannon pseudo-spectral method
\cite{10.1145/365723.365727}
\footnote{Chebyshev collocation methods do not yield symmetric matrices so, for this problem, Fourier collocation methods are advantageous for locally imposing unitarity. Nevertheless, time-symmetric integration with Chebyshev collocation methods still conserve the global U(1) Noether charge, so total probability is conserved regardless of spatial discretization.}.

\begin{figure}
    \includegraphics[scale=0.6]{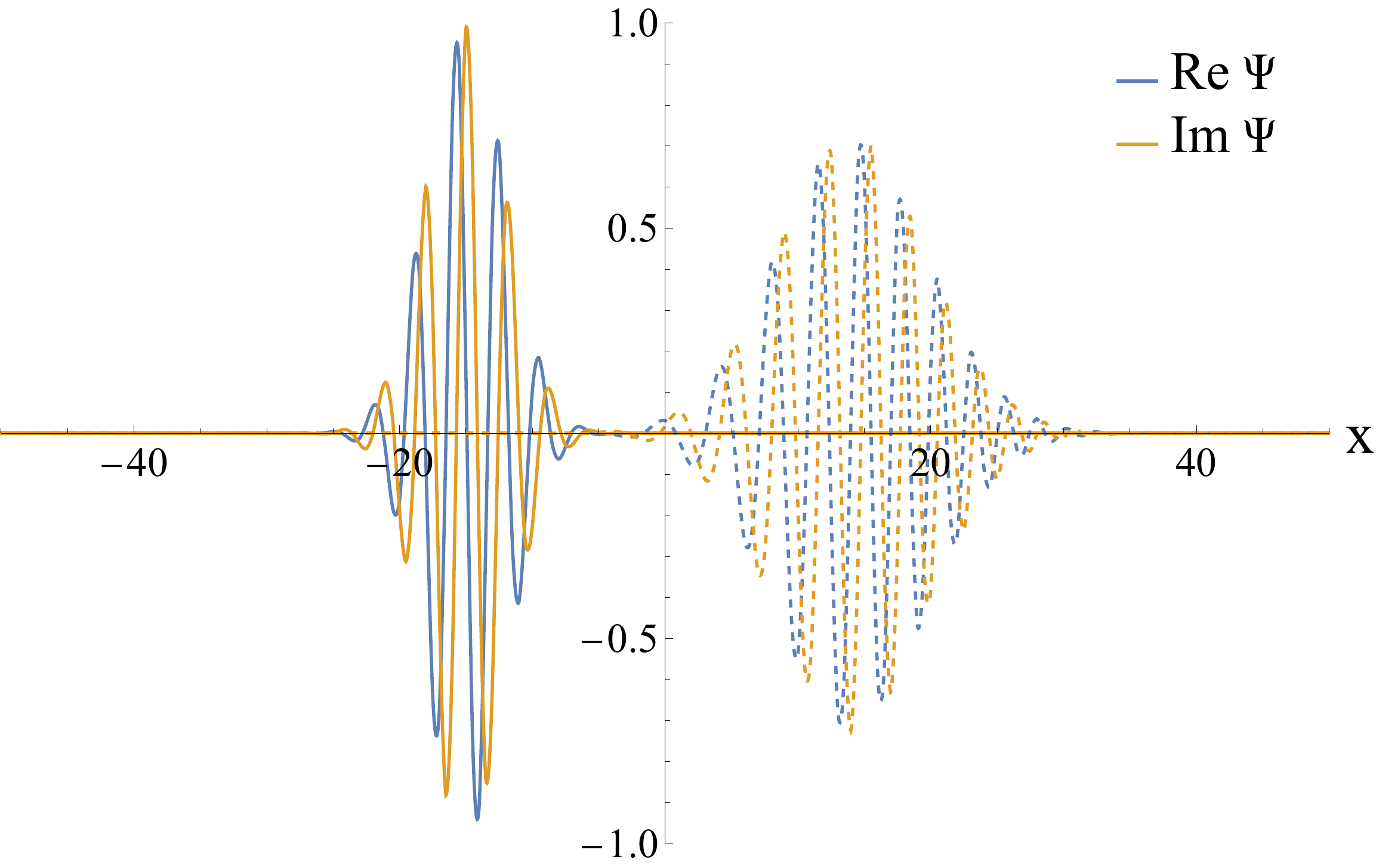}
    \caption{Initial and final states (solid and dashed lines, respectively) for the Schr\"{o}dinger equation evolved with the initial data Eq.~\eqref{eq:SchroInitial} with $w=3$, $k = 2$, and $x_0 = -15$. The scheme H2 with $\Delta t = 0.003$ was used to obtain the final state by evolving from $t=0$ to $t=15$.}
    \label{fig:SchroStates}
\end{figure}

We proceed by selecting the Fourier method and imposing periodic boundary conditions. We use a uniform periodic grid and the symmetric differentiation matrix \eqref{FourierD2}. For our numerical studies, we take the initial data as a Gaussian pulse with width $w$ and momentum $k$ (in units where $\hbar = m = 1$):
\begin{equation}\label{eq:SchroInitial}
    \psi(0,x) = \exp \bigg[ - \frac{(x - x_0)^2}{2 w^2} + {\rm{i}\,} k x \bigg].
\end{equation}
A closed form solution exists for this initial state and is given by
\begin{equation}\label{eq:SchroExact}
    \psi(t,x) = \frac{1}{\sqrt{1 + {\rm{i}\,} t/w^2}} ~ \exp \Bigg\{ \frac{1}{1 + {\rm{i}\,} t/w^2} \bigg[ - \frac{(x - x_0)^2}{2 w^2} + {\rm{i}\,} k \bigg( x - \frac{k t}{2} + {\rm{i}\,} \frac{x_0 t}{w^2} \bigg) \bigg] \Bigg\}
\end{equation}
We use this solution to test the accuracy of  Hermite methods. Using $a =-50$, $b=50$, $N=200$, $w=3$, $k = 2$, and $x_0 = -15$, we evolve the initial data in Eq. \eqref{eq:SchroInitial} from $t=0$ to $t=15$. The initial and final states are shown in Figure \ref{fig:SchroStates}.

\begin{figure}
        \centering
        \begin{subfigure}{0.45\textwidth}
            \centering
            \includegraphics[width=\textwidth]{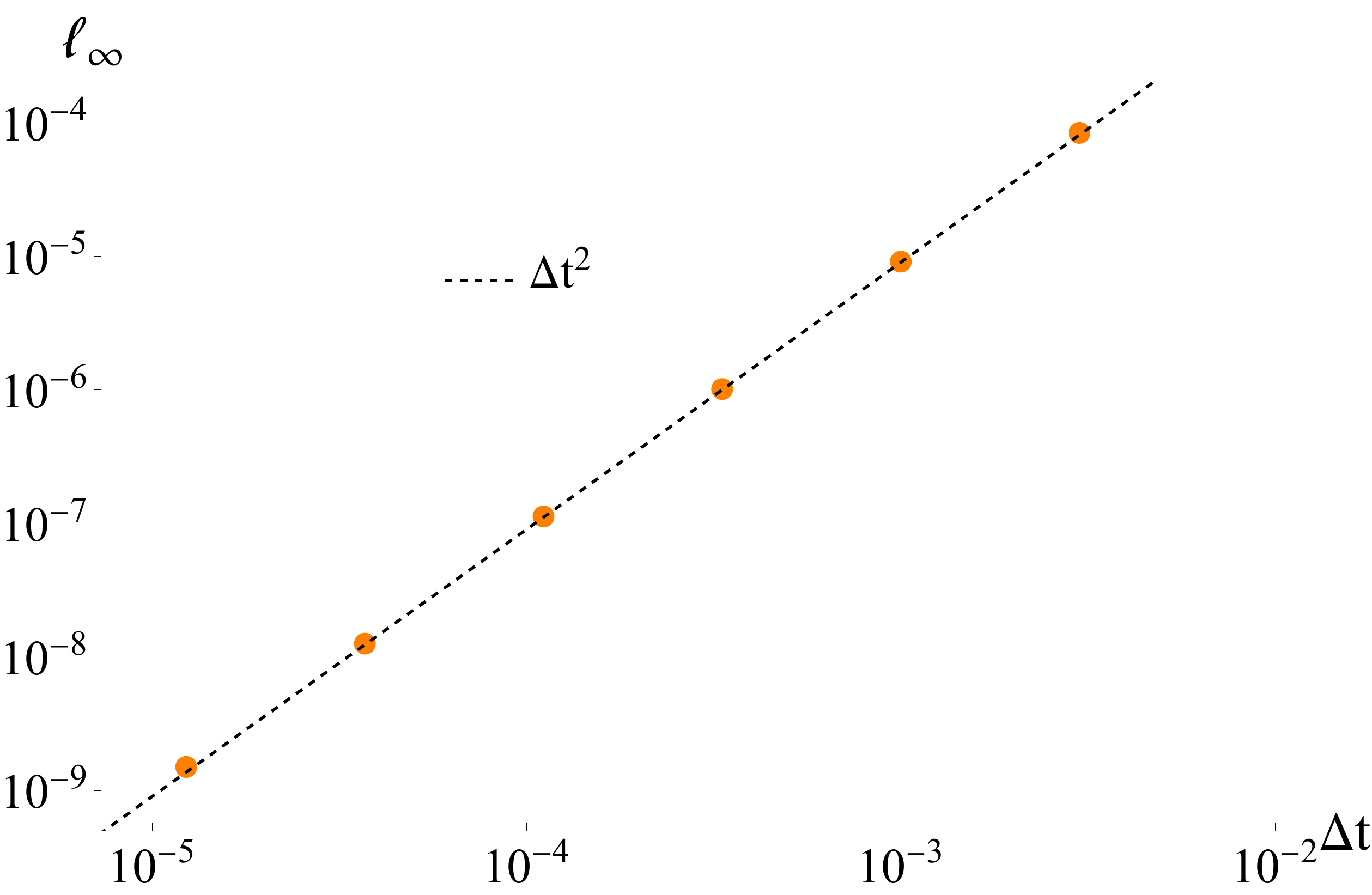}
            \caption{H2 Convergence}
        \end{subfigure}
        \hfill
        \begin{subfigure}{0.45\textwidth}
            \centering
            \includegraphics[width=\textwidth]{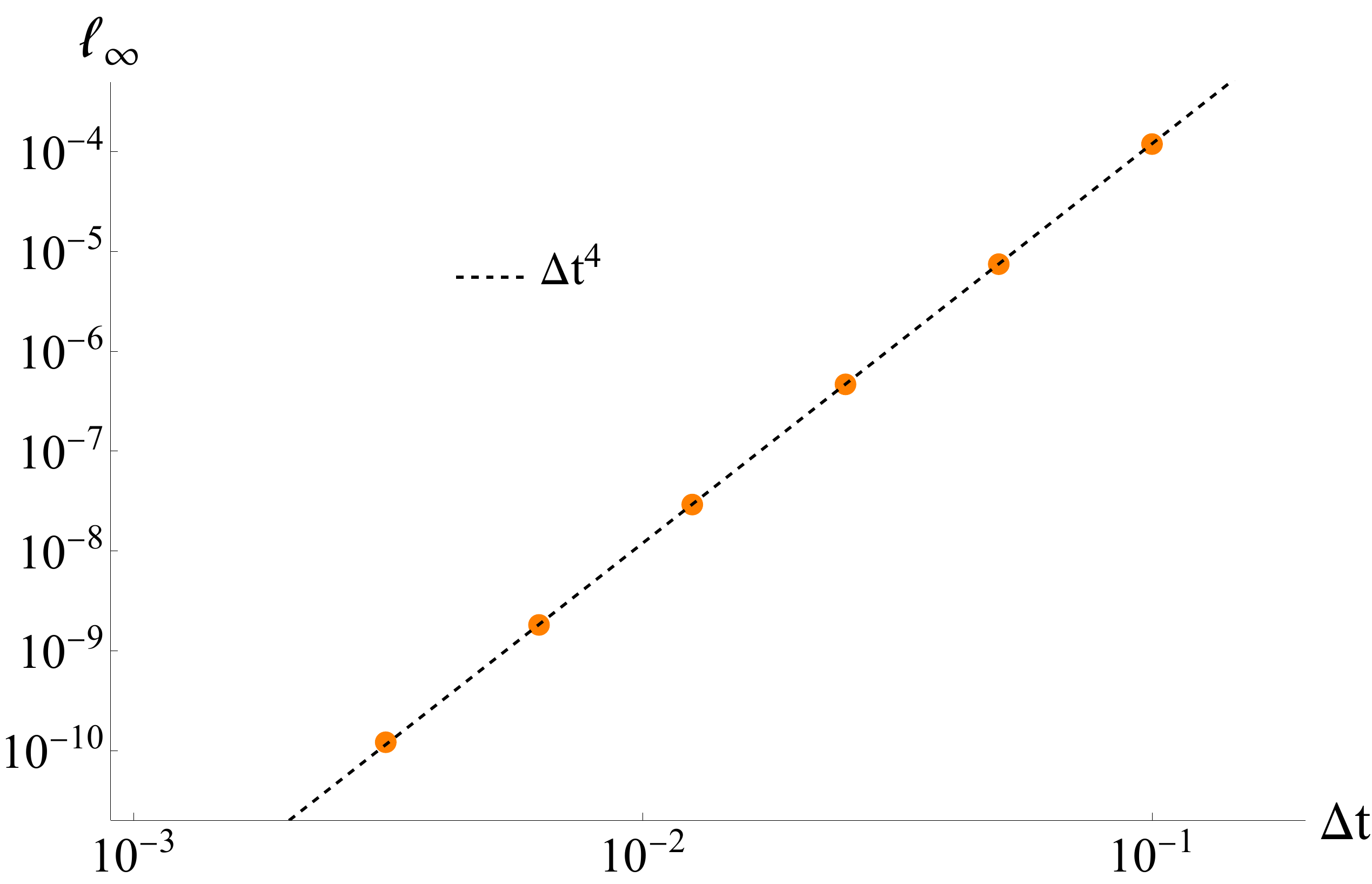}
            \caption{H4 Convergence}
        \end{subfigure}
        \vfill
        \begin{subfigure}{0.45\textwidth}
            \centering
            \includegraphics[width=\textwidth]{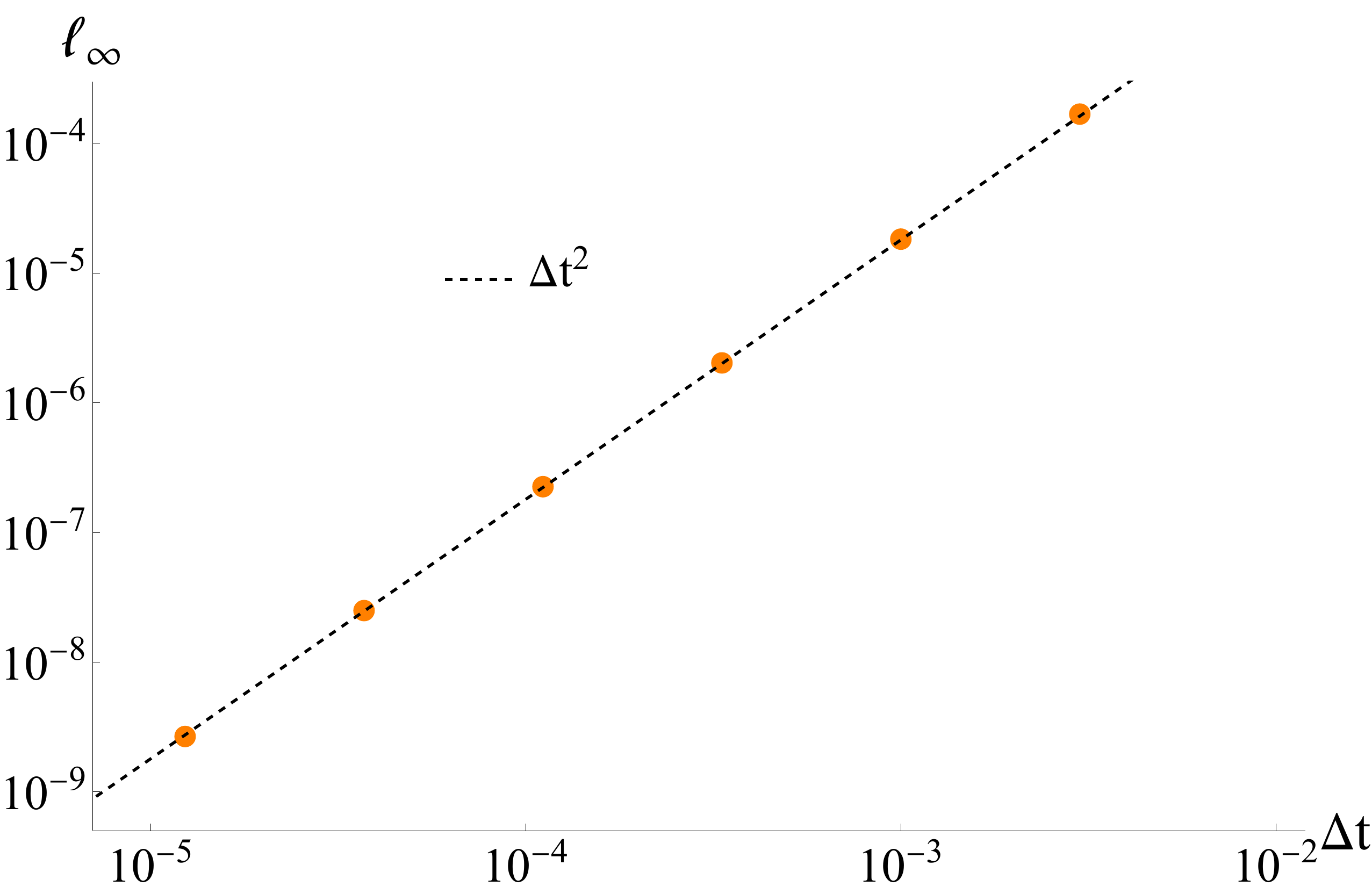}
            \caption{RK2 Convergence}
        \end{subfigure}
        \hfill
        \begin{subfigure}{0.45\textwidth}
            \centering
            \includegraphics[width=\textwidth]{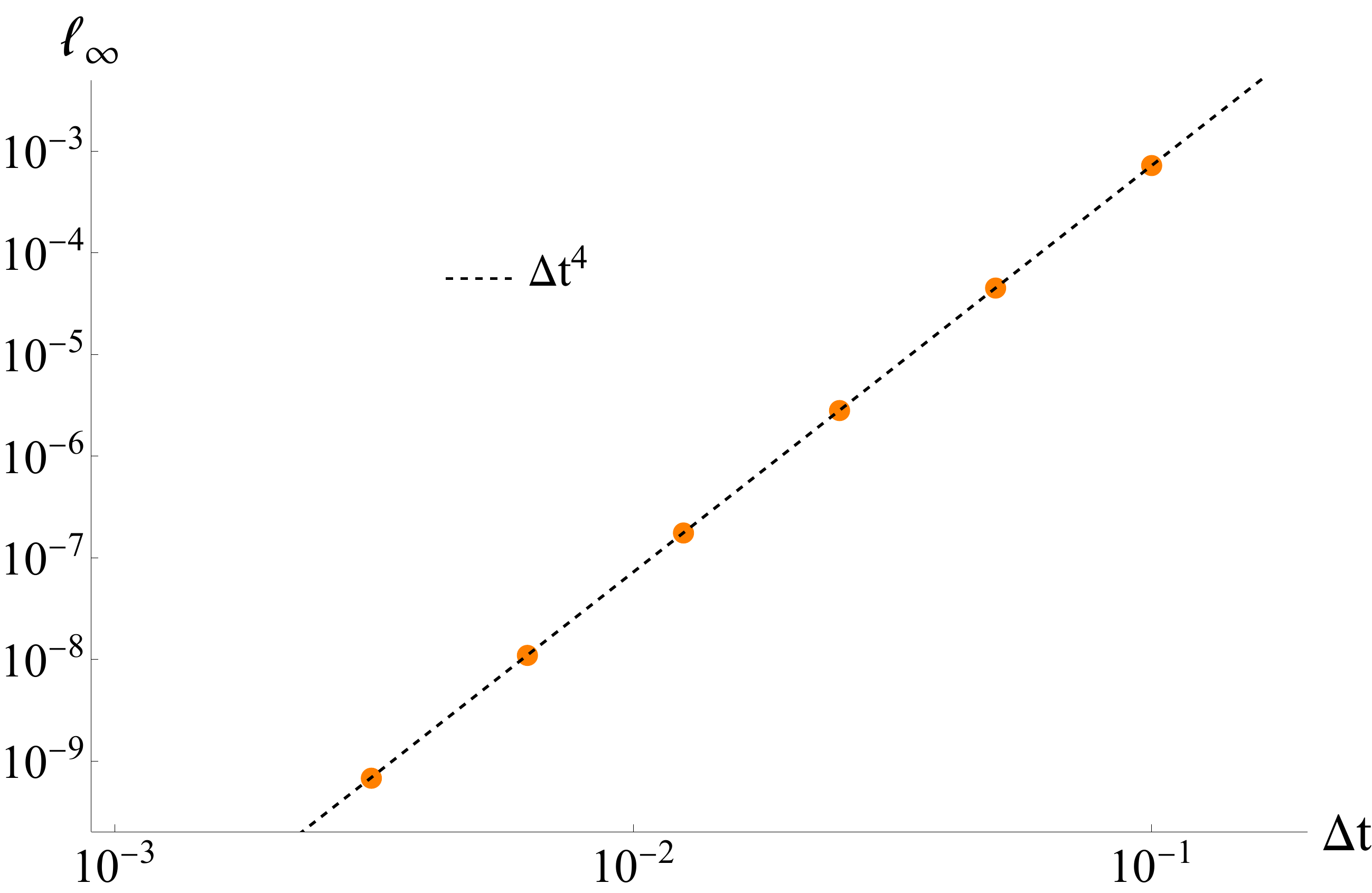}
            \caption{RK4 Convergence}
        \end{subfigure}
        \caption{The $\ell_\infty$ error norm in the evolution of the time-dependent Schr\"{o}dinger equation via the trapezium rule H2 and Hermite rule H4. As expected, the error scales like $\Delta t^2$ and $\Delta t^4$ respectively. For comparison, we also include second- and fourth-order Runge-Kutta methods, whose error also scales like $\Delta t^2$ and $\Delta t^4$ respectively}
        \label{fig:SchroConvergence}
\end{figure}

First, we demonstrate that the schemes H2 and H4 converge by varying the time step size and computing the maximum error between the numerical solution and the exact solution Eq.~\eqref{eq:SchroExact} (the $\ell_\infty$ norm). We show in Figure \ref{fig:SchroConvergence} that the $\ell_\infty$ error norm resulting from the H2 and H4 evolution schemes, given by Eqs.~\eqref{eq:LD2Operator} and \eqref{eq:LD4Operator}, scales like $\Delta t^2$ and $\Delta t^4$ respectively. This is what is predicted by Eq. \eqref{eq:Rerror}.

\begin{figure}
    \centering
    \begin{subfigure}{0.45\textwidth}
        \centering
        \includegraphics[width=\textwidth]{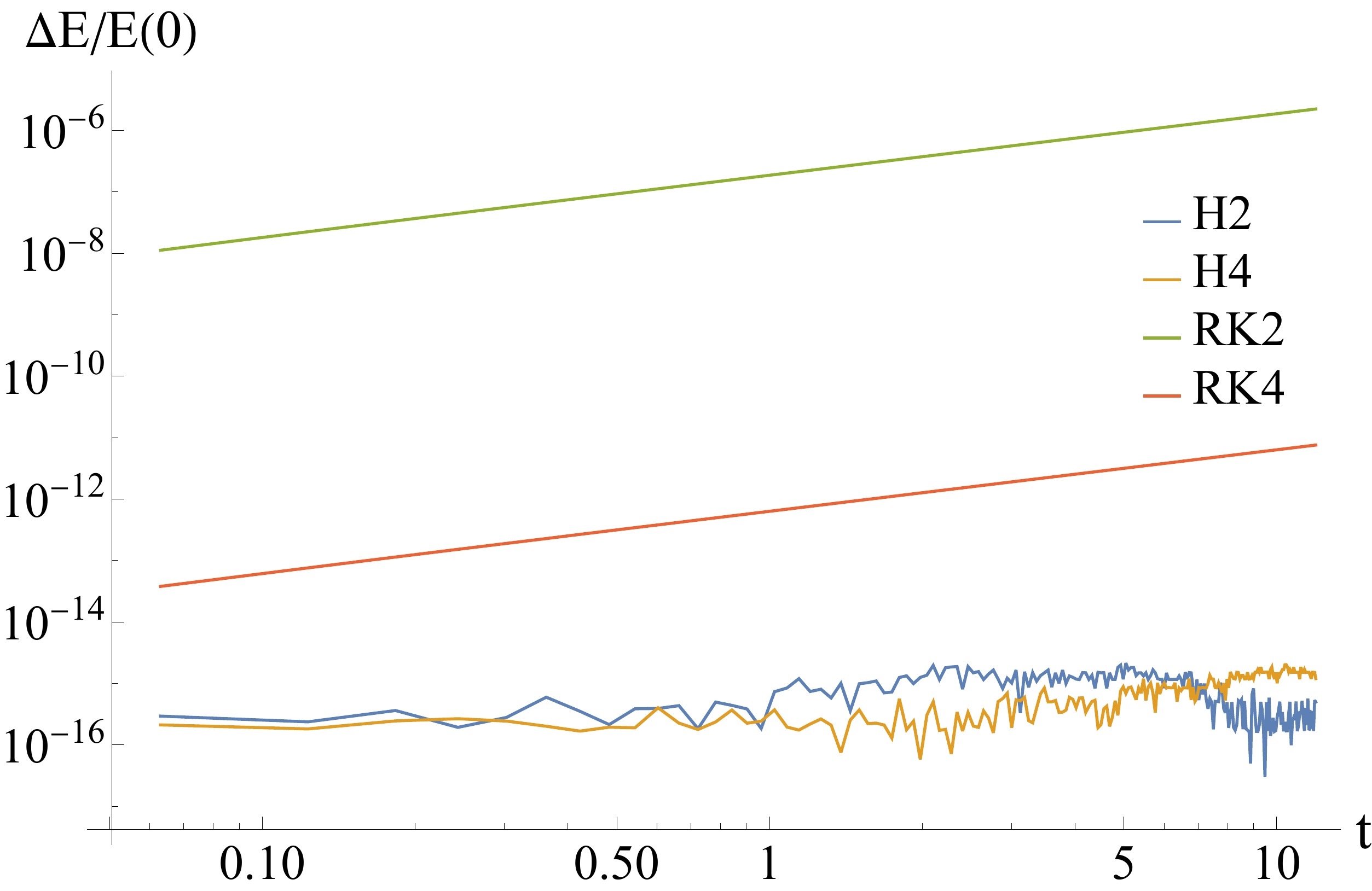}
        \caption{Relative Error in Energy}
    \end{subfigure}
    \hfill
    \begin{subfigure}{0.45\textwidth}
        \centering
        \includegraphics[width=\textwidth]{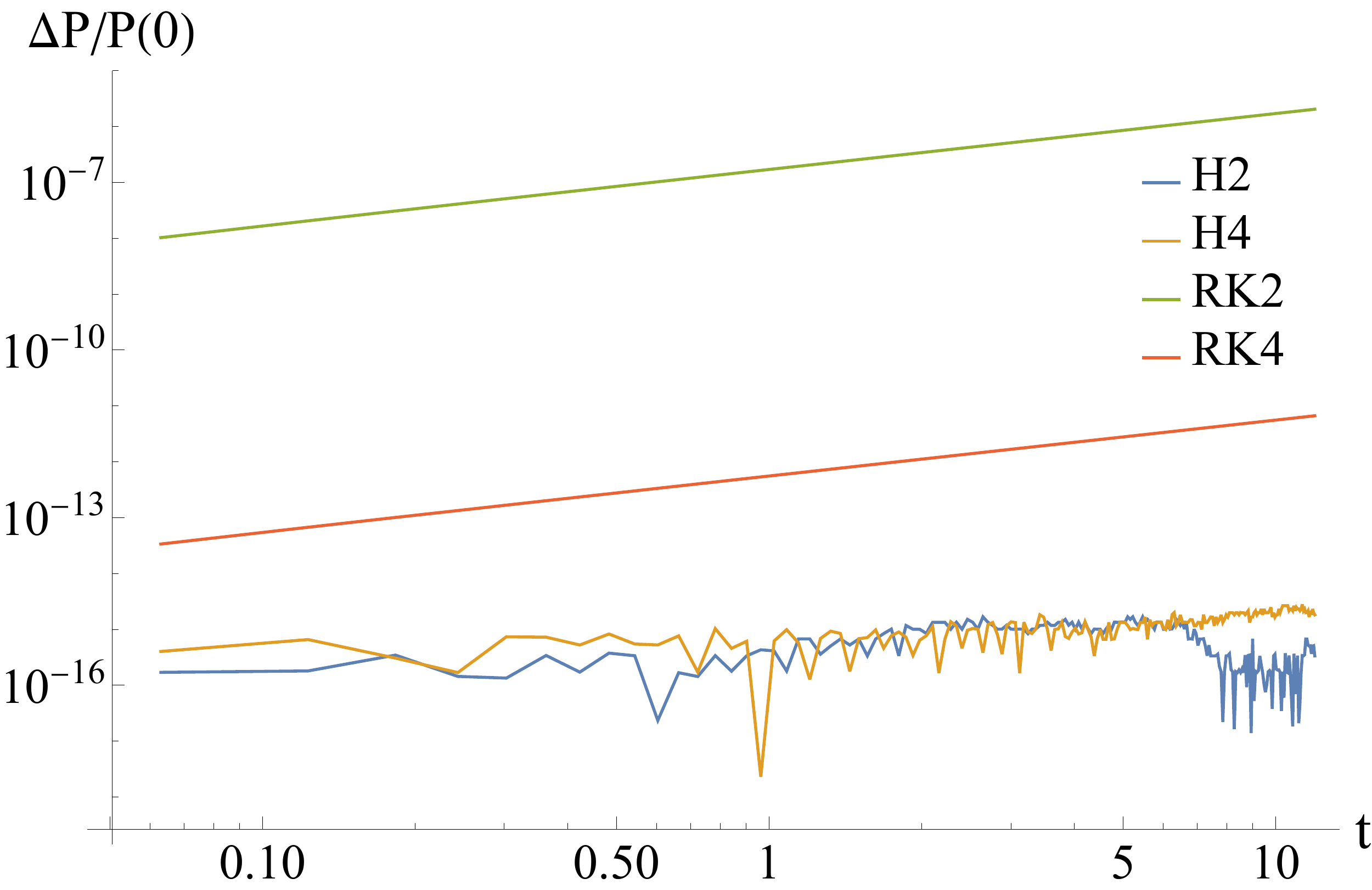}
        \caption{Relative Error in Momentum}
    \end{subfigure}
    \vfill
    \begin{subfigure}{0.45\textwidth}
        \centering
         \includegraphics[width=\textwidth]{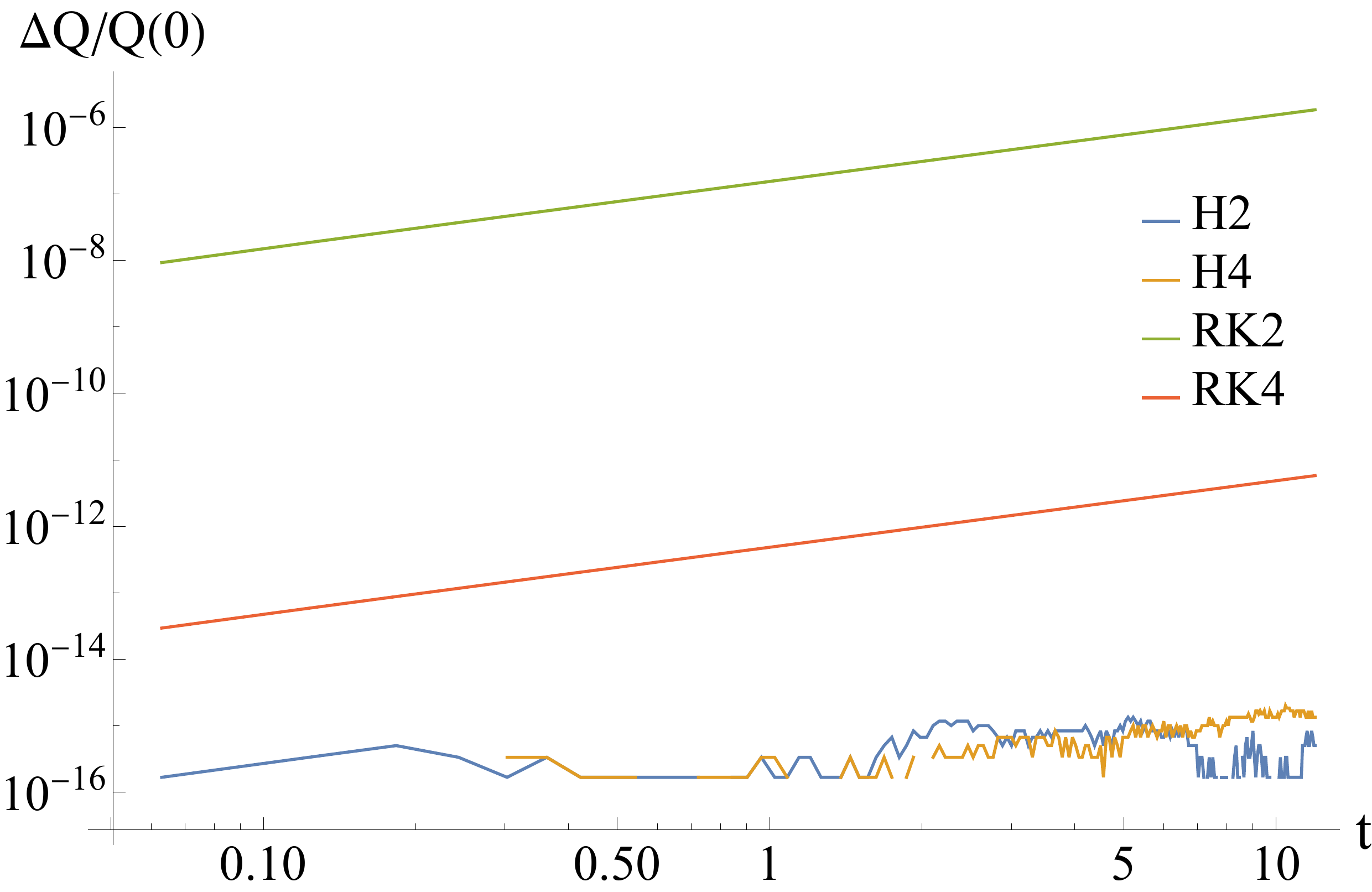}
        \caption{Relative Error in U(1) Charge}
    \end{subfigure}
    \hfill
    \begin{subfigure}{0.45\textwidth}
        \centering
        \includegraphics[width=\textwidth]{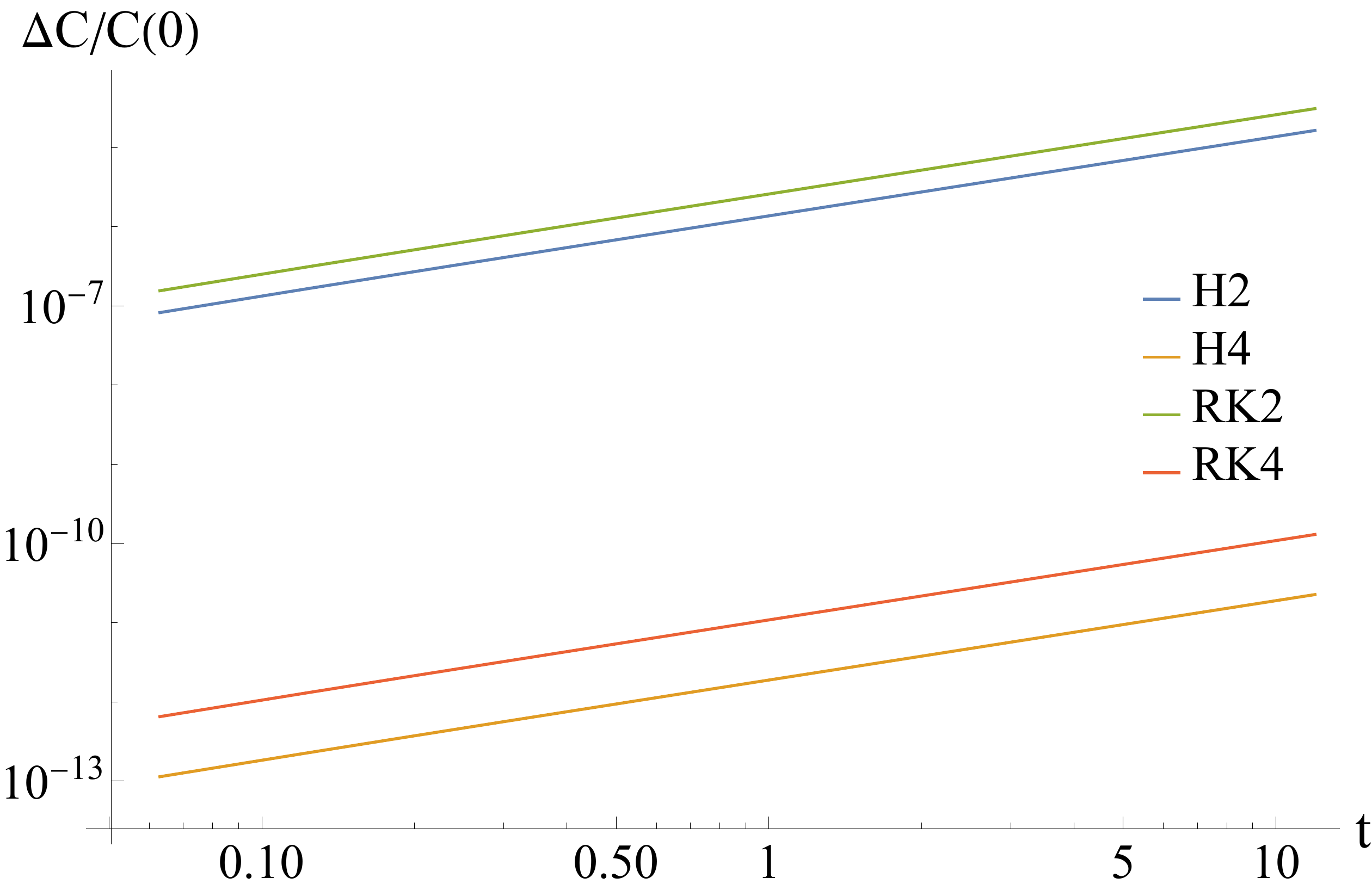}
        \caption{Relative Error in Center-of-Mass Constant}
    \end{subfigure}
    \caption{The relative errors in the Noether charges of the Schr\"odinger equation when numerically evolved. Note that the errors in energy, momentum, and gauge charge remain near machine epsilon when the methods H2 and H4 are used while the error grows without bound when RK2 and RK4 are used. The residual error in H2 and H4 is primarily due to the spatial discretization, particularly due to round-off error entailed by (the non-compensated summations in) matrix-vector multiplications in each time-step. Using compensated summation in matrix-vector multiplications would decrease this error further. While Hermite methods do not conserve the center-of-mass constant, they still outperform Runge-Kutta methods, due to the lower truncation error of the former.}
    \label{fig:SchroConserve}
\end{figure}

Next, we show that certain Noether charges of the action~\eqref{eq:1stOrderAction} are numerically conserved. We discussed four such quantities for a free particle in 1+1 dimensions: a charge related with U(1) gauge symmetry, an energy affiliated with time translation symmetry, a momentum affiliated with space translation symmetry, and a center of mass constant following from Poincar\'e invariance. We evolve the initial data Eq.~\eqref{eq:SchroInitial} with the same parameters as before and a time step of $\Delta t = 0.003$ using Hermite methods while computing the relative error in all four charges as the evolution progresses. As shown in Figure \ref{fig:SchroConserve}, the relative errors of three charges are bounded near machine precision. For comparison, we also evolved this system with explicit RK2 and RK4 schemes, showing that error in these quantities accumulates. (We note that to obtain such low errors for the Hermite methods, it was necessary to calculate the evolution matrices using extended precision, and then round them to double precision before the numerical evolution.) The center of mass constant is not conserved by the Hermite methods, indicating a limitation of the method.

\subsection{Relativistic Scalar Field Theories}\label{sec:SecondOrder}
We now turn to the relativistic scalar field theories of BHPT. We will solve the Klein-Gordon and Regge-Wheeler-Zerilli (RWZ) equations using time-symmetric numerical methods, and demonstrate conservation of Noether-related integrals. Here, we consider the behavior of the respective fields in the spacetime of a non-rotating black hole governed by the Schwarzschild metric (there is no analogue to the RWZ quantities for rotating black holes). Like all spherically symmetric geometries, the Schwarzschild spacetime may be decomposed into the product of two submanifolds, $\mathcal{M}^2 \times S^2$, where $S^2$ is the unit two-sphere and $\mathcal{M}^2$ is spanned by the time and radial coordinates. Notably, this allows the functions defined above to be expanded in scalar spherical harmonics. Thus, the evolution equations for both quantities take the following general form
\begin{equation}\label{eq:GenericCurved}
    \Box \Psi - V_\ell \Psi = 0,
\end{equation}
where $\Box$ is a d'Alembert operator 
\begin{equation}\label{eq:dAlebertian}
    \Box \Psi := \nabla_\alpha \nabla^\alpha \Psi = \frac{1}{\sqrt{-g}} \partial_\alpha (\sqrt{-g} g^{\alpha \beta } \; \partial_\beta \Psi)
\end{equation}
defined on $(\mathcal{M}^2,\mathbf{g})$ and $V_l(x)$ is a time-independent potential unique to each $l$-pole. Unlike the Schr\"{o}dinger equation, Eq.~\eqref{eq:GenericCurved} is second order in time, so application of the Hermite evolution schemes is not as straightforward: it is necessary to perform a first order reduction in time. A common choice is to define a ``momentum density'' by
\begin{equation} \label{eq:nonHam1}
    \Pi := \partial_0 \Psi 
\end{equation}
This allows the evolution equation \eqref{eq:GenericCurved} to be recast in the form
\begin{equation}\label{eq:Generic1p1}
    \partial_0 \Pi = A(x^1) \partial_1 \Pi + B(x^1) \Pi + C(x^1) \partial_1^2 \Psi + E(x^1) \partial_1 \Psi + W(x^1) \Psi.
\end{equation}
The above system is first order in time and second order in space. Upon discretization with the method of lines, the second order spatial derivatives may restrict the Courant limit of the discretized system. In 1+1 dimensions, a first order reduction in space can be facilitated by introducing a new momentum variable 
$\tilde \Pi := \partial_0 \Psi + \lambda(x^1) \partial_1 \Psi$ and selecting a scalar function 
$\lambda(x^1)$ such that the $\partial_1^2 \Psi$ term drops out from Eq.~\eqref{eq:Generic1p1}.
This requirement leads to a quadratic equation, $\lambda^2+A \lambda -C=0$, with roots
 $\lambda=-(\frac{1}{2}A \pm \sqrt{(\frac{1}{2}A)^2+C})$. (These roots coincide with the characteristic speeds of the system). Both of these choices accomplish a first order reduction in space, and one may use this freedom to select the root that yields the least restrictive Courant limit upon discretization, allowing for a stable evolution with larger time-steps. In this work, however, we use time-symmetric methods, which are unconditionally stable, so there is no Courant limit. We will thus make the simplest (and more common) choice $\lambda=0$, which amounts to the original system \eqref{eq:nonHam1} and \eqref{eq:Generic1p1}. 

The Hermite rules can be applied to separately solve each of these equations. Since the system is linear, explicit expressions for $\Psi^{n+1}$ and $\Pi^{n+1}$ may be found. We state these results in Appendix \ref{sec:ExplicitSchemes}, albeit they are rather cumbersome. We will discuss a more streamlined approach now to be used throughout the paper.

We define a state vector
\begin{equation}
    u := 
    \begin{pmatrix}
        \Psi \\
        \Pi
    \end{pmatrix}
\end{equation}
which allows the two first-order equations to be rewritten as a matrix differential equation:
\begin{equation}\label{eq:FirstOrderReduced}
    \partial_0 u = L \,  u
\end{equation}
where
\begin{equation}\label{eq:SecondOrderOps}
    {L} = 
    \begin{pmatrix}
        0 & 1\\
       C  ~ \partial_1^2 + E~ \partial_1 + W & A ~\partial_1 + B
    \end{pmatrix}
\end{equation}
In this form, the expressions obtained in Section~\ref{sec:LinearEquations}, Eqs.~\eqref{eq:LD2Operator} and \eqref{eq:LD4Operator}, may now be applied directly. It is only necessary to specify the quantities $A,B,C,E$ and $W$.
Upon spatial discretization, the differential operator $L$ amounts to a block matrix $\mathbf{L}$ of dimension $(2 N+2) \times (2 N+2)$.

Because all quantities under consideration here share the same form of evolution equation \eqref{eq:GenericCurved}, they may all be derived from a phenomenological $1+1$ action of the form:
\begin{equation}
    S[\Psi,\Psi^\star] =\int_{\mathcal{M}^2} d^2x \; \mathcal{L}=
    - 
    \int_{\mathcal{M}^2} d^2x \sqrt{-g} \Big(g^{\alpha \beta} \nabla_{\alpha} \Psi \nabla_{\beta} \Psi^\star - V_\ell \Psi^\star \Psi \Big)
\end{equation}
The indices $\alpha,\beta=1,2$ label the timelike coordinate $x^0$ and the radial spacelike coordinate $x^1$; the metric $g_{\alpha \beta}$ and its determinant $g$ are defined on $\mathcal{M}^2$.

With this action functional, it is now a simple matter to determine conserved currents and charges analogous to those of the previous section. The U(1) gauge symmetry $\Psi \rightarrow e^{{\rm{i} \,}  \delta \alpha} \Psi$ is Noether-related to the conserved current
\begin{equation}
    J^\alpha = {\rm{i} \,} ( \Psi^\star \nabla^\alpha \Psi - \Psi \nabla^\alpha \Psi^\star )
\end{equation}
and the conserved global charge
\begin{equation}\label{eq:GenericCharge}
    Q = \int_{\Sigma_0} dx^1 \sqrt{-g} ~ J^0 = {\rm{i} \,} \int_{\Sigma_0} dx^1 \sqrt{-g} ~ g^{0 \beta} (\Psi^\star \partial_\beta \Psi - \Psi \partial_\beta \Psi^\star ).
\end{equation}

We follow Poisson \cite{Poisson:2009pwt} in defining the canonical momentum by
\begin{equation} \label{eq:canmom}
    \Pi = \frac{\partial \mathcal{L}}{\partial (\partial_0 \Psi^\star)} = \sqrt{-g} ~ g^{0 \beta} \partial_\beta \Psi
\end{equation}
and a Hamiltonian density by
\begin{equation*}
    \mathcal{H} = \Pi^\star \partial_0 \Psi + \Pi \partial_0 \Psi^\star - \mathcal{L}
\end{equation*}
\begin{equation}
    = - \frac{1}{2 g^{0 0}} \bigg( \frac{\Pi^\star \Pi}{\sqrt{-g}} + g^{0 1} (\Pi^\star \partial_1 \Psi + \Pi \partial_1 \Psi^\star) \bigg) + \frac{\sqrt{-g}}{2} \Bigg( \bigg(g^{1 1} - \frac{(g^{0 1})^2}{g^{0 0}} \bigg) \partial_1 \Psi^\star \partial_1 \Psi + V_\ell \Psi^\star \Psi \Bigg) 
\end{equation}
leading to canonical equations of motion,
\begin{equation} \label{eq:Ham1}
    \partial_0 \Psi = - \frac{1}{g^{0 0}} \bigg( \frac{\Pi}{\sqrt{-g}} + g^{0 1} \partial_1 \Psi \bigg)
\end{equation}
\begin{equation} \label{eq:Ham2}
    \partial_0 \Pi = - \sqrt{-g} V_\ell \Psi + \partial_1 \Bigg( \sqrt{-g} \bigg(g^{1 1} - \frac{(g^{0 1})^2}{g^{0 0}} \bigg) \partial_1 \Psi - \frac{g^{0 1}}{g^{0 0}} \Pi \Bigg),
\end{equation}
and a conserved symplectic form \cite{Crnkovic:1986ex,Prabhu:2018jvy,Green:2019nam}.
Implementing the canonical equations \eqref{eq:Ham1}-\eqref{eq:Ham2} leads to a matrix L that differs from \eqref{eq:SecondOrderOps}. When the system is discretized in time using an explicit method, the Hamiltonian approach may have a different Courant limit. One may again choose to subtract a term $\partial_1 \Psi$ from the canonical momentum to perform a first order reduction in space, but we leave this approach for future work.

The canonical momentum 
\eqref{eq:canmom}
typically appears in first-order symmetric hyperbolic formulations of the Klein-Gordon equation. In these formulations, the matrix $L$ appearing in Eq.~\eqref{eq:FirstOrderReduced} 
is symmetric and positive definite. We do not pursue a fully first-order formulation here, for three reasons: \textit{(i)} For 1+1 systems, a fully first order reduction of the type discussed earlier (solving an algebraic quadratic equation to eliminate second spatial derivatives from the system) is possible in the Schwarzschild spacetime (whence spherical harmonic modes of the Bardeen-Press-Teukolsky equation are uncoupled, as dicussed below), but is not as straightforward in Kerr spacetime (due to the fact that the 1+1 Teukolsky equation exhibits mode coupling, one must solve a large matrix quadratic equation to eliminate second spatial derivatives from the system). \textit{(ii)} One may alternatively achieve a first order reduction in space by evolving the spatial gradient of $\Psi$ separately, but this increases the number of variables, and introduces a constraint that may be violated numerically. \textit{(iii)} A second order formulation in space and first order in time leads to a Hamiltonian that, upon discretization with the method of lines, is analogous to the Hamiltonian of a system of coupled harmonic oscillators. The latter approach makes it straightforward to establish conservation of energy and symplectic structure. We have thus opted for a formulation first order in time and second order in space in this work. (Nevertheless, the methods outlined here are applicable to a fully first order system as well.)


Spacetime symmetries (diffeomorphisms) are Noether-related to components of the stress-energy tensor
\begin{equation}\label{eq:RelativisticStress}
    T^\alpha_{\;\;\;\;\beta} = 2 g^{\alpha \gamma} \nabla_{(\gamma} \Psi \nabla_{\beta)} \Psi^\star - g^\alpha_{\;\;\;\;\beta} \big( g^{\gamma \delta}\nabla_{\gamma} \Psi \nabla_{\delta} \Psi^\star - V_\ell \Psi^\star \Psi \big).
\end{equation}
where index parentheses denote symmetrization. In particular, if the vector field $k^\alpha$ is Lie-derives the metric $g_{\alpha \beta}$ and the potential $V_l$, then $k^\beta T^\alpha_{\;\;\;\;\beta}$ is a conserved Noether current.

If neither $g_{\alpha \beta}$ nor $V_\ell$ have explicit $x^0$ dependence, then the energy
\begin{equation}\label{eq:RelativisticEnergy}
    E =\int_{\Sigma_0} dx^1 \mathcal{H} =\int_{\Sigma_0} dx^1 \sqrt{-g} T_{\;\;\;\;0}^0 =\int_{\Sigma_0} dx^1 \sqrt{-g} \Big( g^{00} \partial_{0} \Psi \partial_{0} \Psi^\star - g^{1 1} \partial_{1} \Psi \partial_{1} \Psi^\star + V_\ell \Psi^\star \Psi \Big),
\end{equation}
is a constant of evolution.

For homogeneous wave equations (i.e. without particle sources), the canonical equations~\eqref{eq:Ham1}-\eqref{eq:Ham2}
are preferable over the non-canonical equations
\eqref{eq:nonHam1}-\eqref{eq:Generic1p1}
when explicit (e.g. Runge-Kutta) methods are used for time evolution, because the canonical equations admit a Courant factor (and thus CFL limit on the time step 
$\Delta t$) up to an order of magnitude higher compared to the non-canonical equations.
Nevertheless, this work is based on implicit (Hermite) integration schemes, which are unconditionally stable (i.e. CFL unlimited), and thus we opted to use the non-canonical equations 
\eqref{eq:nonHam1}-\eqref{eq:Generic1p1}
for our numerical implementation.

\subsubsection{Klein-Gordon Field in Flat Spacetime}
As a first example, we consider the massless Klein-Gordon equation in flat 1+1 dimensional spacetime. That is, we take $g_{\alpha \beta} = \eta_{\alpha \beta}$, so 
\begin{equation}\label{eq:MinkMet}
    ds^2 = \eta_{\alpha \beta}dx^\alpha dx^\beta = -dt^2 + dx^2,
\end{equation}
which is the line element for Minkowski space in the usual Cartesian coordinates. In these coodrinates, the Klein-Gordon field obeys the classical wave equation:
\begin{equation}\label{eq:FlatWave}
    -\partial_t^2 \Psi +\partial_x^2 \Psi = 0.
\end{equation}
This is a special case of the above considerations if $\mathcal{M}^2$ is taken as Minkowski space rather than the Schwarzschild submanifold and $V_l = 0$. We consider this case because it admits closed form solutions, allowing the accuracy of our numerical schemes to be tested exactly. If the coordinates are taken as the standard Cartesian spacetime coordinates of Minkowski space $t,x$, then setting $C=1$ and $A=B=E=V=0$ in the operator \eqref{eq:SecondOrderOps} yields
\begin{equation}
    {L} = 
    \begin{pmatrix}
        0 & 1\\
       \partial_x^2  & 0
    \end{pmatrix}.
\end{equation}
In addition, the exact solution to the flat wave equation~\eqref{eq:FlatWave} is $\Psi(t,x) = f(x - t) ~ + ~ g(x + t)$, where $f$ and $g$ are arbitrary smooth functions.

\begin{figure}
    \centering
    \begin{subfigure}{0.45\textwidth}
        \centering
        \includegraphics[width=\textwidth]{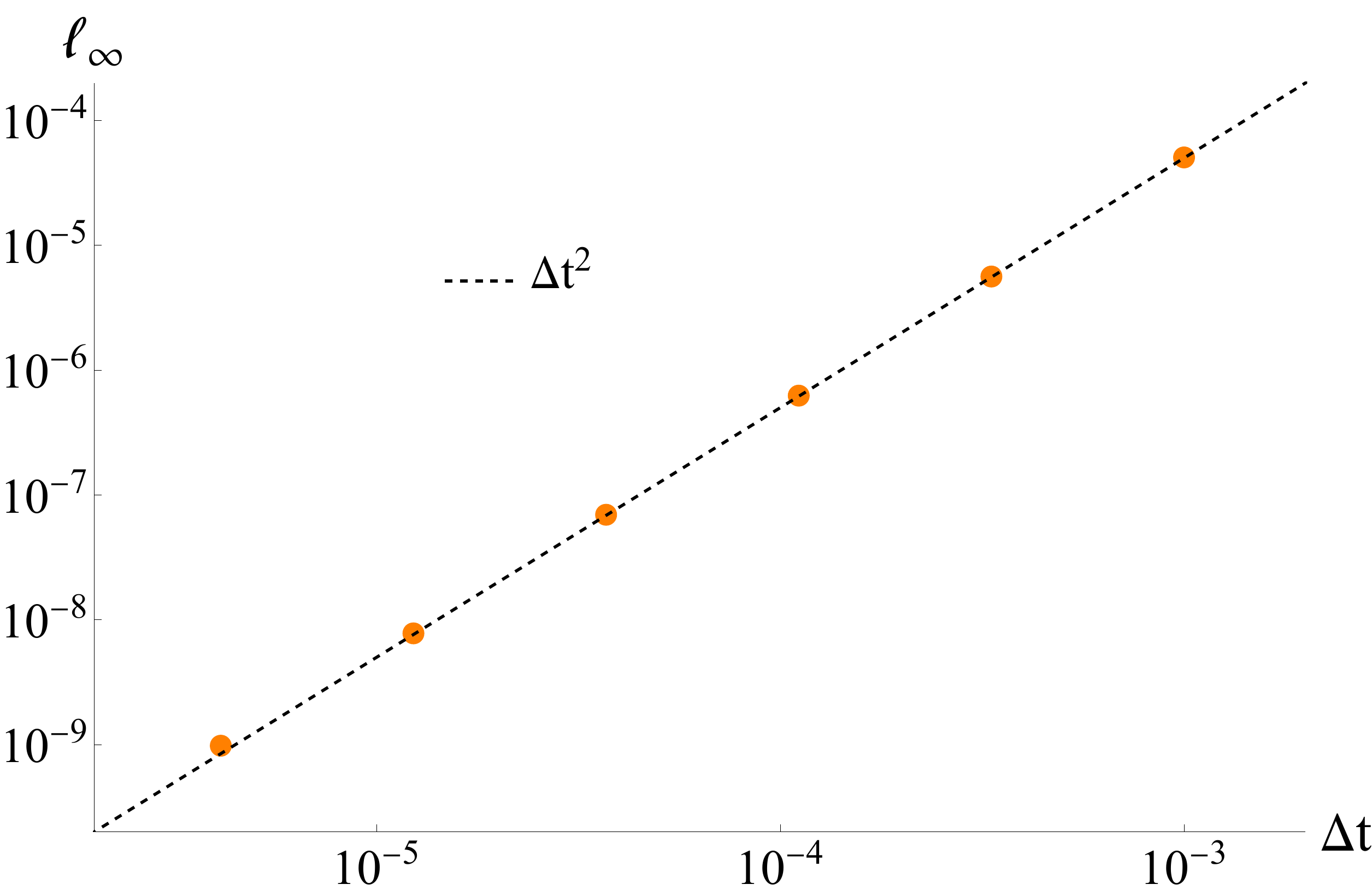}
        \caption{H2 Convergence}
    \end{subfigure}
    \hfill
    \begin{subfigure}{0.45\textwidth}
        \centering
        \includegraphics[width=\textwidth]{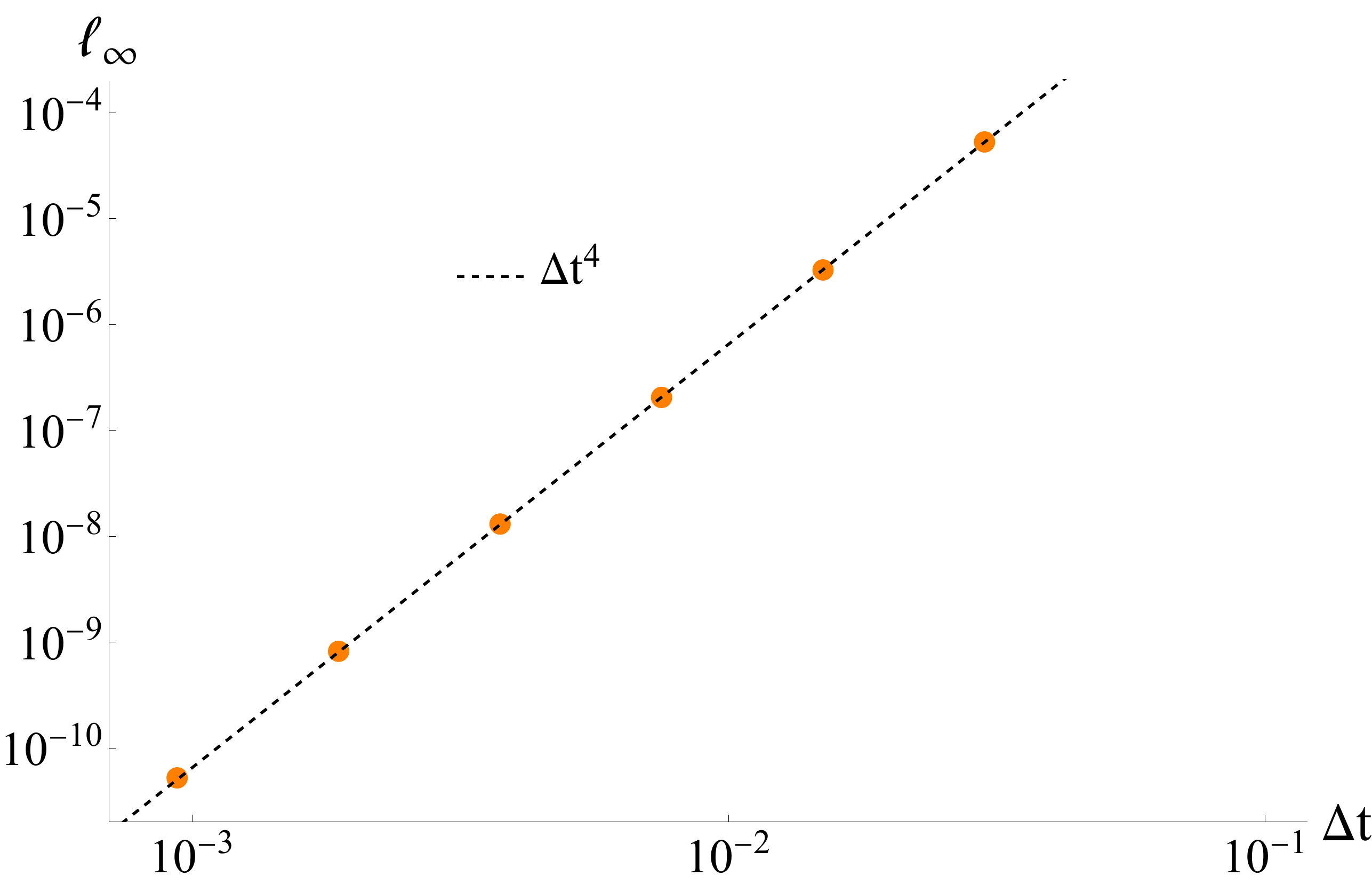}
        \caption{H4 Convergence}
    \end{subfigure}
    \caption{The $\ell_\infty$ error norm in the evolution of the time-dependent Schr\"{o}dinger equation via the trapezium rule H2 and Hermite rule H4. As with the Schr\"{o}dinger equation, the error scales like $\Delta t^2$ and $\Delta t^4$ respectively.}
    \label{fig:ClassicalConvergence}
\end{figure}

For our numerical studies, we consider the Cauchy initial data
\begin{equation}
    \Psi(0,x) = \exp \bigg[ - \frac{(x - x_0)^2}{w^2} \bigg], \quad \Pi(0,x) = \frac{2 (x - x_0)}{w^2} \exp \bigg[ - \frac{(x - x_0)^2}{w^2} \bigg]
\end{equation}
which gives rise to the exact solution
\begin{equation}\label{eq:ClassicalExact}
    \Psi(t,x) = \exp \bigg[ - \frac{(x - x_0 - t)^2}{w^2} \bigg].
\end{equation}
We take $x_0 = - 2.8$ and $w = 1/3$. We also use a Fourier pseudo-spectral method  (Eqs.~\eqref{EquidistantNodes} - \eqref{FourierD2}) for spatial discretization in this problem, taking $a=-5$, $b=5$, and $N=200$. Note that this imposes periodic boundary conditions. We first vary the time steps used in the methods H2 and H4 to verify that they converge. This is shown in Figure \ref{fig:ClassicalConvergence}, where the $\ell_\infty$ error of the numerical solution compared to the exact solution Eq.~\eqref{eq:ClassicalExact} is plotted against the time step used in evolution.

Next, we consider the Noether charges associated with this field. The U(1) gauge charge is trivial since we are considering a real field, and we have already demonstrated that the center-of-mass constant is not conserved by Hermite methods. However, we do examine the energy defined by Eq.~\eqref{eq:RelativisticEnergy} and we show that Hermite methods conserve it to machine precision. In addition, in this example, $V_\ell = 0$ and the metric $g_{a b}$ does not depend on the spatial coordinate $x^1$, so the charge associated with the $b=1$ Noether current in Eq. \eqref{eq:RelativisticStress}, the momentum, is also a constant of evolution:
\begin{equation}
    P = \int_{\Sigma_t} dx ~ \partial_t \Psi \partial_x \Psi
\end{equation}
We find that Hermite methods conserve this quantity as well.

\subsubsection{Symplectic structure}
The Klein-Gordon equation \eqref{eq:FlatWave}, upon first-order in time reduction and spatial discretization on a grid $\mathbf{X}=\{x_i\}_{i=0}^N$ with the method of lines, can be written as a system of ordinary differential equations, in the discretized canonical form:
\begin{subequations} \label{DiscreteHam0}
\begin{eqnarray} \label{DiscreteHam1}
& &\frac{{d\bf{\Psi} }}{{dt}} = \frac{{\partial H}}{{\partial {\bf{\Pi}}}}={\bf{\Pi}} \\ \label{DiscreteHam2}
& &\frac{{d{\bf{\Pi}}}}{{dt}} = -\frac{{\partial H}}{{\partial {\bf{\Psi}}}}= {{\bf{D}}_2} \cdot \bf{\Psi} 
\end{eqnarray}
\end{subequations}
where
\begin{equation} \label{DiscreteHam}
H({\bf{\Psi}},{\bf{\Pi}})= \frac{1}{2}{{\bf{\Pi}}^{\rm{T}}} \cdot {\bf{\Pi}} - \frac{1}{2}{({{\bf{D}}^{(2)}} \cdot \bf{\Psi}  )^{\rm{T}}} \cdot ({{\bf{D}}^{(2)}}  \cdot \bf{\Psi}  )
\end{equation}
is the Hamiltonian, $\mathbf{\Pi}=\{\Pi_i\}_{i=0}^N$ denotes the canonical momenta, $\mathbf{\Psi}=\{\Psi_i\}_{i=0}^N$ their conjugate variables,
the supescript $\text{T}$ denotes matrix transpose and 
$\cdot$ denotes the dot (or inner) product of two matrices. The Hamiltonian $H$ has a similar form to that of a system of coupled harmonic oscillators, with the differentiation matrices responsible for the coupling.

If a time-stepping scheme is a canonical transformation, then the infinitesimal phase-space volume:
\begin{equation}
 d \mathbf{\Psi}^{n+1} d \mathbf{\Pi}^{n+1} =  d \mathbf{\Psi}^{n} d \mathbf{\Pi}^{n} 
\end{equation}
is conserved from one time-step $t_n$ to the next, $t_{n+1}$,
in agreement with Liouville's theorem. Equivalently, if a time-step amounts to a canonical transformation, then the Jacobian of the transformation must be equal to unity:
\begin{equation} \label{Jacobian}
J = \frac{{\partial (\mathbf{\Psi}^{n+1},\mathbf{\Pi}^{n+1})}}{{\partial (\mathbf{\Psi}^{n},\mathbf{\Pi}^{n})}} = 1.
\end{equation}
A Runge-Kutta method, or, equivalently, a 1-point Taylor expansion integration rule \eqref{eq:Taylorint} applied to the system \eqref{DiscreteHam0} leads to a Jacobian that deviates from unity in each time-step; this violation is monotonic and accumulates over time. As a result, these methods violate Liouville's theorem and are not symplectic, and thus are unsuitable for precise long-time evolutions of Hamiltonian systems \cite{markakis_time-symmetry_2019}. In contrast, with a time-symmetric method, such as Hermite integration or, equivalently, a 2-point Taylor expansion
\eqref{eq:GeneralHermiteRule} applied to the system \eqref{DiscreteHam0}, it can easily be shown (cf.~Appendix~\ref{sec:ExplicitSchemes}) that the Jacobian \eqref{Jacobian} remains exactly equal to unity at all times. That is, Hermite integration methods are volume preserving for this system. This property can be shown to hold for any quadratic Hamiltonian, that is, for all linearized equations arising in black-hole perturbation theory. This makes them an excellent method for long-time numerical evolution and gravitational-wave extraction from EMRI simulations.

\begin{figure}
    \centering
    \begin{subfigure}{0.45\textwidth}
        \centering
        \includegraphics[width=\textwidth]{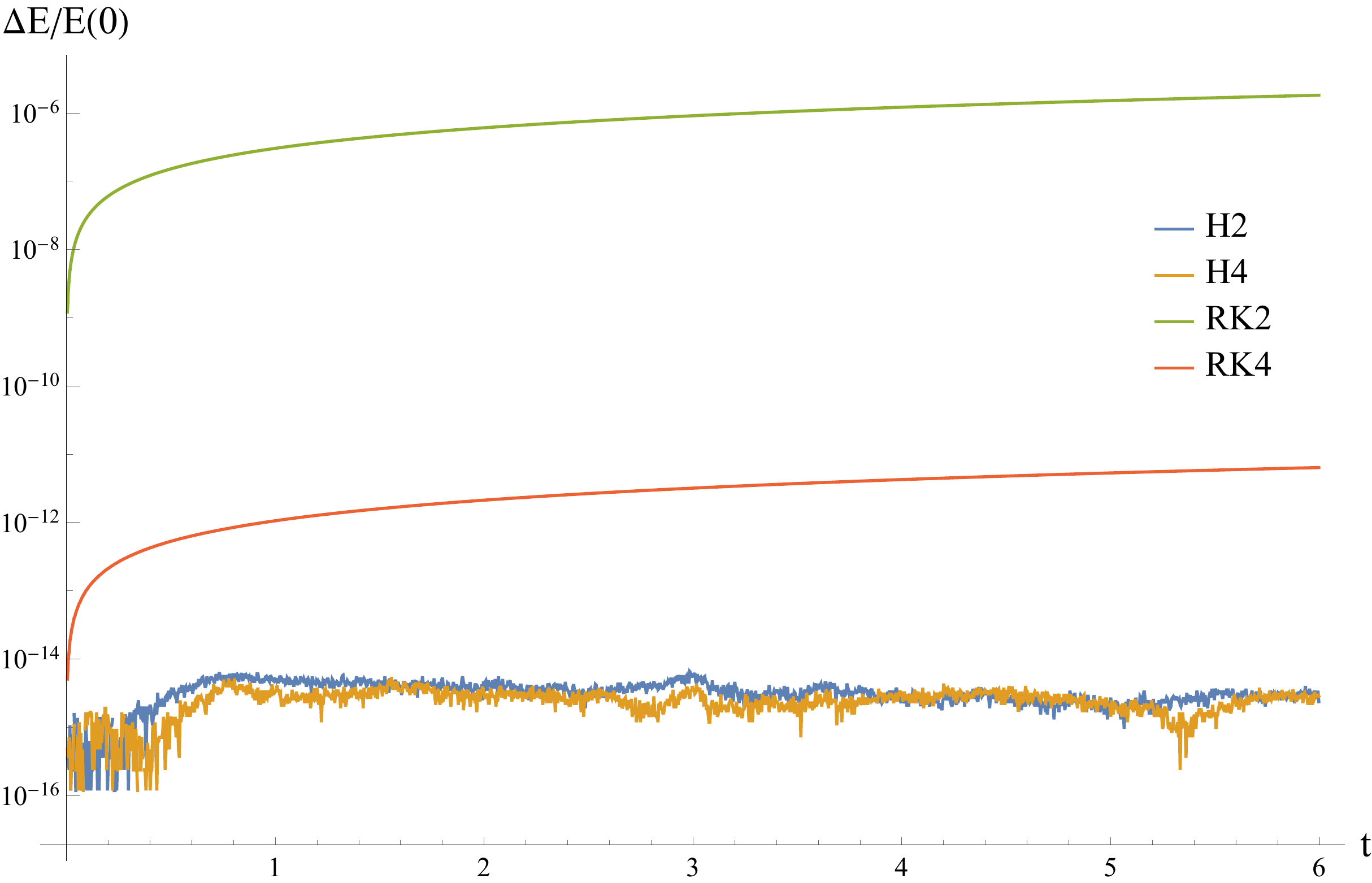}
        \caption{Relative Error in Energy}
    \end{subfigure}
    \hfill
    \begin{subfigure}{0.45\textwidth}
        \centering
        \includegraphics[width=\textwidth]{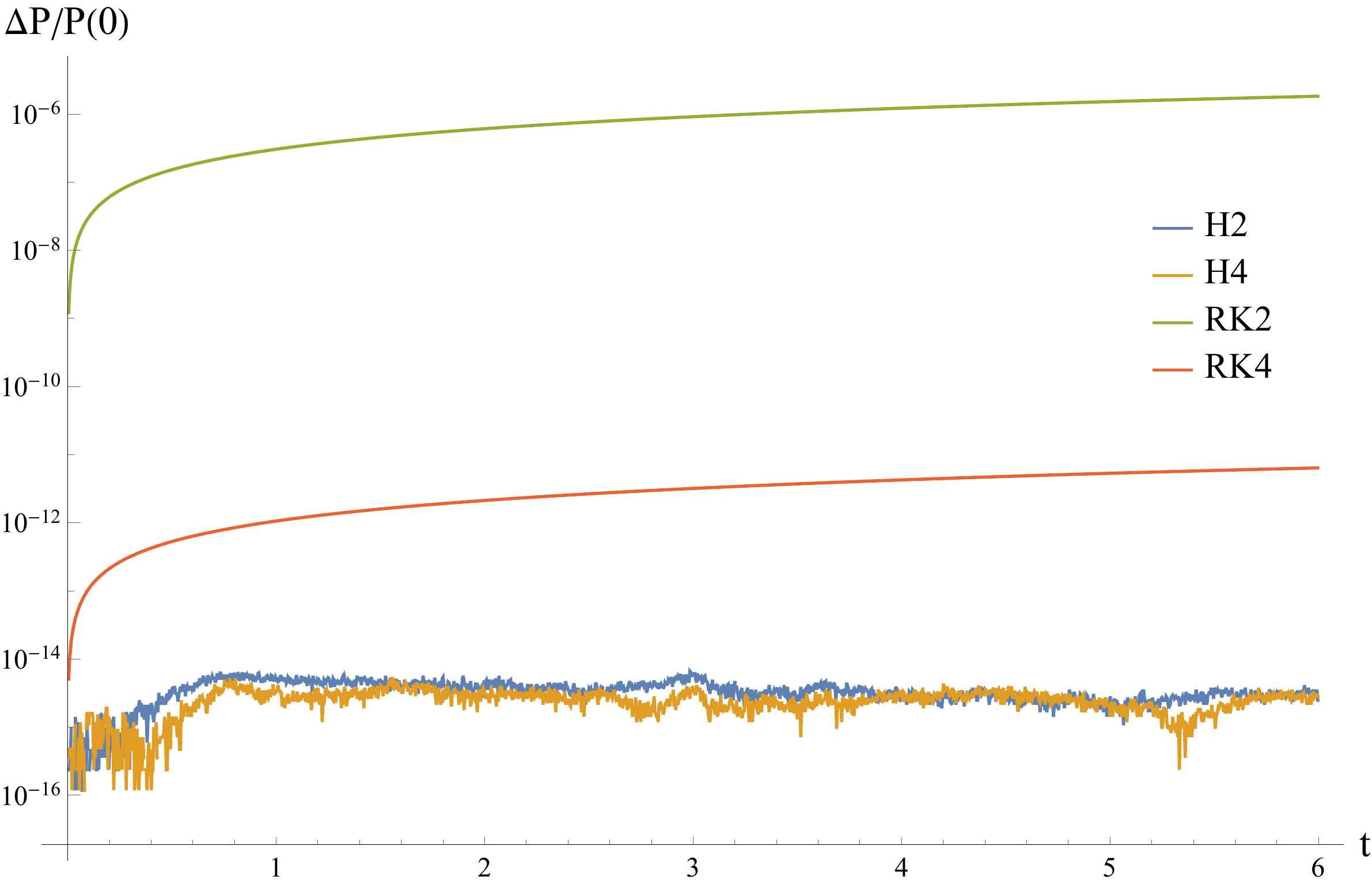}
        \caption{Relative Error in Momentum}
    \end{subfigure}
    \caption{The relative errors in the Noether charges of the flat-space Klein-Gordon equation when numerically evolved. Note that the errors in energy and momentum remain near machine epsilon when the methods H2 and H4 are used while the error grows without bound when RK2 and RK4 are used. As before, the residual error in H2 and H4 is primarily due to the spatial discretization.}
    \label{fig:ClassicalConserve}
\end{figure}

\subsubsection{Klein-Gordon Field in Schwarzschild Spacetime}\label{sec:BHScalar}
Next, we consider the masslesss Klein-Gordon equation outside a non-rotating black hole. Unlike the previous problems, we will now impose astrophysical boundary conditions on this field. This amounts to the requirement no new information can enter the computational domain from either the black hole horizon or from null infinity (the surface future pointing null rays tend towards if they exit the black hole spacetime). A very convenient way to do this is to choose a hyperboloidally compact coordinate system. That is, coordinates where $r=\infty$ is located at a finite spatial coordinate and where time slices intersect null infinity as $r \rightarrow \infty$ rather than spacelike infinity \cite{Zenginoglu:2008wc,zenginoglu_hyperboloidal_2008,Zenginoglu:2008uc,zenginoglu_geometric_2011,Gautam:2021ilg}. There are several choices that achieve this \cite{zenginoglu_hyperboloidal_2008,racz_numerical_2011}, but we find that the ``minimal gauge'' defined by Ansorg and Macedo \cite{ansorg_spectral_2016} yields the simplest algebraic expressions and covers the entire black hole exterior with a single hyperboloidal layer. 
Starting from the Schwarzschild metric in ingoing (horizon penetrating) Eddington-Finkelstein coordinates $\{v,r,\theta,\phi\}$,
\begin{equation} \label{eq:Eddington-Finkelstein}
    ds^2 = - \bigg( 1 - \frac{2 M}{r} \bigg) dv^2 + 2 dv dr + r^2 d\Omega^2,
\end{equation}
where $d\Omega^2=d\theta^2+\sin^2 \theta d \phi^2$ is the metric on the unit 2-sphere $S^2$, $v$ is a null coordinate and $r$ is the Schwarzschild (areal) radial coordinate,
we perform the tortoise coordinate transformation
\begin{subequations}\label{eq:Tortoisext} 
\begin{eqnarray}  
     \frac{dx}{dr} \mkern-12mu &=& \mkern-12mu \frac{1}{{4M}}
     \left(1 - \frac{{2M}}{r}\right)^{-1},
     \label{eq:Tortoisex} \\
    t \mkern-12mu &= & \mkern-12mu \frac{v}{4M}-x. \label{eq:Tortoiset} 
\end{eqnarray}
\end{subequations}
The Schwarzschild metric in tortoise coordinates $\{t,x,\theta,\phi\}$ is given by the line element 
\begin{equation}\label{eq:TortoiseMetric}
d{s^2} = 16{M^2}\left(1 - \frac{{2M}}{r}\right)( - d{t^2} + d{x^2}) + {r^2}d{\Omega ^2}.
\end{equation}
We now seek a hyperboloidal transformation
\begin{subequations} \label{eq:Hyperboloidal}
\begin{eqnarray}
t \mkern-12mu  &=& \mkern-12mu \tau - h(\sigma) \label{eq:Hyperboloidalt}\\
x \mkern-12mu  &=& \mkern-12mu  g(\sigma) \label{eq:Hyperboloidax}
\end{eqnarray}
\end{subequations}
where
\begin{eqnarray} \label{eq:sigma}
    \sigma := \frac{2 M}{r}.
\end{eqnarray}
is a compactified radial coordinate, such that $\sigma([2M,\infty]) \rightarrow [0,1]$, with future null infinity ${\mathscr{I}^+}$ located at $\sigma_{\mathscr{I}^+} = 0$ and the black hole event horizon ${\mathscr{H}^+}$ located at $\sigma_{\mathscr{H}^+} = 1$.
Integrating Eq.~\eqref{eq:Tortoisex} yields
\begin{equation}  \label{eq:gofsigma}
g(\sigma)=\frac{1}{2} \left[\frac{1}{\sigma}-\ln \sigma
+\ln(1-\sigma) \right].
\end{equation}
The height function $h(\sigma)$ may be obtained by asymptotically integrating outgoing null rays\footnote{Alternatively, one may substitute Eqs.~\eqref{eq:Hyperboloidal} and \eqref{eq:gofsigma} into Eq.~\eqref{eq:TortoiseMetric} and require regularity on the boundaries to obtain $h(\sigma)$ \cite{test1}.} \cite{Schinkel:2013tka}. Eq.~\eqref{eq:Eddington-Finkelstein}, for  $ds=0, \;\; d\theta=0, \;\; d\phi=0, \;\; dv dr >0$, yields
\begin{equation} 
dv=
\frac{2dr}{1-\frac{2M}{r}}=2 \Bigg(1 + \frac{{2M}}{r} + \frac{{4{M^2}}}{{{r^2}}} + O({r^{ - 3}})\Bigg)dr 
\nonumber
\end{equation}
 which is integrated to
\begin{equation} 
\frac{v}{{4M}} = {\rm{constant}} + \frac{r}{{2M}} + \ln \frac{r}{{2M}} +
\frac{{2M}}{r} + O({r^{ - 2}}). 
\nonumber
\end{equation}
The ``minimal gauge'' can be imposed by truncating this asymptotic expansion to next-to-leading order\cite{Schinkel:2013tka,ansorg_spectral_2016}.  By virtue of Eq.~\eqref{eq:sigma}, this motivates the ansatz
\begin{equation}  \label{eq:vtransform} 
\frac{v}{{4M}} = \tau +  \frac{1}{\sigma} - \ln \sigma. 
\end{equation}
In the new chart $\{\tau, \sigma, \theta, \phi\}$, the compactified hyperboloidal slices $\Sigma_\tau$ are described by $\tau = \rm{constant}$. 
Substituting Eq.~\eqref{eq:vtransform} into Eq.~\eqref{eq:Tortoiset} yields Eq.~\eqref{eq:Hyperboloidalt} with
\begin{equation}  \label{eq:hofx} 
h(x)=\frac{1}{2} \left[-\frac{1}{\sigma}+\ln \sigma
+\ln(1-\sigma)
\right].
\end{equation}
Substituting the transformations \eqref{eq:Hyperboloidal} into
\eqref{eq:TortoiseMetric} yields the Schwarzschild metric in hyperboloidal coordinates: 
\begin{equation}\label{eq:HyperboloidalMetric}
d{s^2} = 16{M^2}(1 - \sigma )\Bigg[ - d{\tau ^2} + \frac{{1 - 2{\sigma ^2}}}{{{\sigma ^2}(1 - \sigma )}}d\tau d\sigma  + \frac{{1 + \sigma }}{{{\sigma ^2}(1 - \sigma )}}d{\sigma ^2} + \frac{1}{{4{\sigma ^2}(1 - \sigma )}}d{\Omega ^2}\Bigg].
\end{equation}
The free (massless) Klein-Gordon equation~\eqref{eq:GenericCurved} 
in these coordinates 
is singular at $\sigma = 0$. To obtain a regular equation, suitable for numerical evolution, we define a new quantity $\varphi := \Psi / \sigma$. We can thus arrive at a $1+1$ formulation by expanding the field in spherical harmonics,
\begin{equation}
    \Psi(\tau,\sigma,\theta,\phi) = \sum_{\ell=0}^\infty ~ \sum_{m = -\ell}^\ell \sigma ~\varphi_{\ell } (\tau, \sigma) Y_{\ell m}(\theta,\phi).
\end{equation}
The resulting equation for $\varphi_{\ell }$ is regular and reads
\begin{multline}\label{eq:LittlePsiEqn}
    -(1 + \sigma) \partial_\tau^2 \varphi + (1-2\sigma^2) \partial_\sigma \partial_\tau \varphi  + (1-\sigma) \sigma^2 \partial_\sigma^2 \varphi\\
    - 2 \sigma \partial_\tau \varphi + \sigma(2-3\sigma) \partial_\sigma \varphi - (\ell ( \ell + 1) + \sigma) \varphi = 0,
\end{multline}
where we dropped the subscript $\ell$ for brevity. The above equation is polynomial in $\sigma$ and regular at the boundaries.
As alluded to earlier, we have covered the black whole exterior with a single hyperboloidal chart, in order to \textit{(i)} automatically impose outflow boundary conditions on the event horizon ${\mathscr{H}^+}$ and future null infinity ${\mathscr{I}^+}$, \textit{(ii)} compactify the infinite domain to a finite computational domain, \textit{(iii)} avoid multiple hyperboloidal layers that can introduce unneccessary code complexity and numerical artifacts near multi-domain boundaries, and \textit{(iv)} extract gravitational waves at ${\mathscr{I}^+}$ (which is now included in the computational domain).
A hyperbolicity analysis confirms that the above equation is strongly hyperbolic, that the outgoing characteristic speed $\lambda_{\rm{out}}=\sigma-1 \le 0$ vanishes on the event horizon ($\sigma_{\mathscr{H}^+}=1 $), and the incoming characteristic speed $\lambda_{\rm{in}}=\frac{\sigma^2}{1+\sigma} \ge 0$ vanishes at future null infinity ($\sigma_{\mathscr{I}^+}=0$). This ensures that, upon discretization with the method of lines, the correct boundary conditions -- no incoming waves at ${\mathscr{I}^+}$ and no outgoing waves on ${\mathscr{H}^+}$-- will be satisfied automatically, regardless of spatial discretization scheme. This is a valuable property, as it means that the boundary conditions will be automatically embedded in our differential operators (which, upon discretization, will amount to matrices), and the Hermite schemes outlined in the previous section are readily applicable without any modification at the boundaries.

The above equation, takes the form of Eq.~\eqref{eq:GenericCurved} on a flat submanifold $\mathcal{N}^2$ spanned by $\tau$ and $\sigma$, with line element,
\begin{equation} \label{eq:hyperbMink}
    \eta_{\alpha \beta}dx^\alpha dx^\beta =-dt^2+dx^2= - d\tau^2 + \frac{1 - 2 \sigma^2}{\sigma^2 (1-\sigma)} ~ d\tau d \sigma + \frac{1+\sigma}{\sigma^2(1-\sigma)} d\sigma^2.
\end{equation}
That is, Eq.~\eqref{eq:LittlePsiEqn} can be written in the 1+1 covariant form
\begin{equation}\label{eq:LittlePhiEqnCovar}
\eta^{\alpha \beta} \nabla_{\alpha} \nabla_{\beta} \varphi - V_\ell \;  \varphi  = 0,
\end{equation}
where $V_\ell = 4 \sigma^2(1-\sigma)(\ell(\ell+1) +\sigma)$ is an effective potential and $\eta_{\alpha \beta}$ is an effective metric \cite{jaramillo_pseudospectrum_2021} given by Eq.~\eqref{eq:hyperbMink}. 
Eq.~\eqref{eq:LittlePhiEqnCovar} stems from the flat-metric action\footnote{It is of course possible to begin from the standard action for the Klein-Gordon field using the original, physical Schwarzschild metric and arrive at the same field equation.},
\begin{equation}\label{eq:ActionEffective}
    S[\varphi,\varphi^\star] =
    - \frac{1}{2} \int_{\mathcal{N}^2} d^2 x \sqrt{-\eta} \Big( \eta^{\alpha \beta} \nabla_{\alpha} \varphi^\star \nabla_{\beta} \varphi + V_\ell \; \varphi^\star \varphi \Big).
\end{equation}
Extremizing this action functional with respect to $\varphi^\star$ and substituting the Minkowksi metric in Cartesian coordinates \eqref{eq:MinkMet} yields the Klein-Gordon equation
$[\partial_t^2-\partial_x^2+V_\ell (x)] \varphi$.
The RWZ equations also stem directly from this action for a different effective potential (cf.~Appendix~\ref{sec:RWZ}). However, since implementation of boundary conditions in Schwarzschild or tortoise coordinates is computationally complicated, the hyperboloidal coordinates \eqref{eq:hyperbMink} will also be used to numerically evolve the RWZ equations and BPT equations below.
Extremizing this action functional and substituting the Minkowksi metric in hyperboloidal coordinates \eqref{eq:hyperbMink} yields 
Eq.~\eqref{eq:LittlePsiEqn}. 

Noether-related conserved quantities immediately follow for each $(\ell,m)$-mode from the action functional~\eqref{eq:ActionEffective}. As demonstrated earlier, invariance with respect to U(1) gauge tranformations and time translations is Noether-related to th conserved charge 
\begin{equation}\label{eq:KGGauge}
    Q = \int_{\Sigma_\tau} d\sigma \Big[ 2 (1+\sigma) (\varphi^\star \partial_\tau \varphi - \varphi ~\partial_\tau \varphi^\star ) + (1-2\sigma^2) (\varphi^\star \partial_\sigma \varphi - \varphi~ \partial_\sigma \varphi^\star ) \Big]
\end{equation}
and energy
\begin{equation}\label{eq:KGEnergy}
    E = \int_{\Sigma_\tau} d\sigma \Big[ (1+\sigma) \partial_\tau \varphi^\star \partial_\tau \varphi + \sigma^2 (1-\sigma) \partial_\sigma \varphi^\star \partial_\sigma \varphi + (\ell(\ell+1) + \sigma) \varphi^\star \varphi \Big].
\end{equation}
These expressions follow from the covariant expressions \eqref{eq:GenericCharge} and \eqref{eq:RelativisticEnergy},  specialized to the chart \eqref{eq:hyperbMink}.
For our numerical studies, we use the momentum variable $\Pi = \partial_\tau \varphi$ to reduce 
the Klein-Gordon equation \eqref{eq:LittlePsiEqn} to the first-order in time form of Eq.~\eqref{eq:FirstOrderReduced}, with the evolution operator $L$
given by Eq.~\eqref{eq:SecondOrderOps}, with
\begin{equation}\label{eq:ScalarL}
    A(\sigma) = \frac{1 - 2 \sigma^2}{1+\sigma}, \quad B(\sigma) = - \frac{2 \sigma}{1+\sigma}
\end{equation}
\begin{equation}\label{eq:ScalarM}
    C(\sigma) = \sigma^2 \frac{1 - \sigma}{1 + \sigma}, \quad E(\sigma) = \sigma \frac{2-3\sigma}{1+\sigma}, \quad W(\sigma) = -\frac{\ell(\ell+1) + \sigma}{1+\sigma}.
\end{equation}

In a method of lines framework, the fields $\varphi(\tau,x)$, $\Pi(\tau,\sigma)$ are evaluated on a discrete spatial grid $\boldsymbol {\sigma}=\{\sigma_\imath\}_{\imath=0}^N$ so that $\varphi(\tau,\sigma) \rightarrow \boldsymbol {\varphi} (\tau) $ and $\Pi(\tau,\sigma) \rightarrow \boldsymbol {\Pi} (\tau) $. The components $\varphi(\tau,\sigma_\imath) := \varphi_\imath(\tau)$ and 
 $\Pi(\tau,\sigma_\imath) := \Pi_\imath(\tau)$
of the vectors $\boldsymbol{\varphi} (\tau)$ and $\boldsymbol{\Pi} (\tau)$ are the values of the fields evaluated on the gridpoints. 
Then, Eq.~\eqref{eq:LittlePsiEqn} heuristically amounts to a system of $2N+2$ ODEs of the form 
\eqref{eq:FirstOrderReduced} in one time variable $\tau$:
\begin{equation} \label{eq:phipiODE}
\frac{d}{{d{\tau}}}
    \begin{pmatrix}
        \varphi_\imath \\
        \Pi_\imath
    \end{pmatrix}
= \sum\limits_{\jmath = 0}^N {{\begin{pmatrix}
0&{{\delta _{\imath \jmath}}}\\
{{C_\imath }D_{\imath \jmath}^{(2)} + {E_\imath}D_{\imath \jmath}^{(1)} + {W_\imath }{\delta _{\imath \jmath}}}&{{A_\imath }D_{\imath \jmath}^{(1)} + {B_\imath }{\delta _{\imath \jmath}}}
\end{pmatrix}} 
    \begin{pmatrix}
        \varphi_\jmath \\
        \Pi_\jmath
    \end{pmatrix}
},\quad \imath = 0,1,...,N
\end{equation}
where 
$A_\imath =A(\sigma_\imath )$, 
$B_\imath =B(\sigma_\imath )$, 
$C_\imath =C(\sigma_\imath )$, 
$E_\imath =E(\sigma_\imath )$, 
$W_\imath =W(\sigma_\imath )$,
$\delta _{\imath \jmath}$ is the Kronecker delta,
and no summation over $\imath$ is implied.
The outflow boundary conditions imposed through hyperboloidal slicing preclude a Fourier pseudo-spectral method (which is associated with periodic boundary conditions). We instead use the Chebyshev pseudo-spectral method (Eqs. \eqref{ChebyshevNodes} - \eqref{ChebyshevD2}), on the interval $\sigma \in [0,1]$ with $N=200$ nodes. The system \eqref{eq:phipiODE} is then evolved
via the Runge-Kutta or Hermite schemes outlined in 
Sec.~\ref{Hermite}.

\begin{figure}
    \centering
    \begin{subfigure}{0.45\textwidth}
        \centering
         \includegraphics[width=\textwidth]{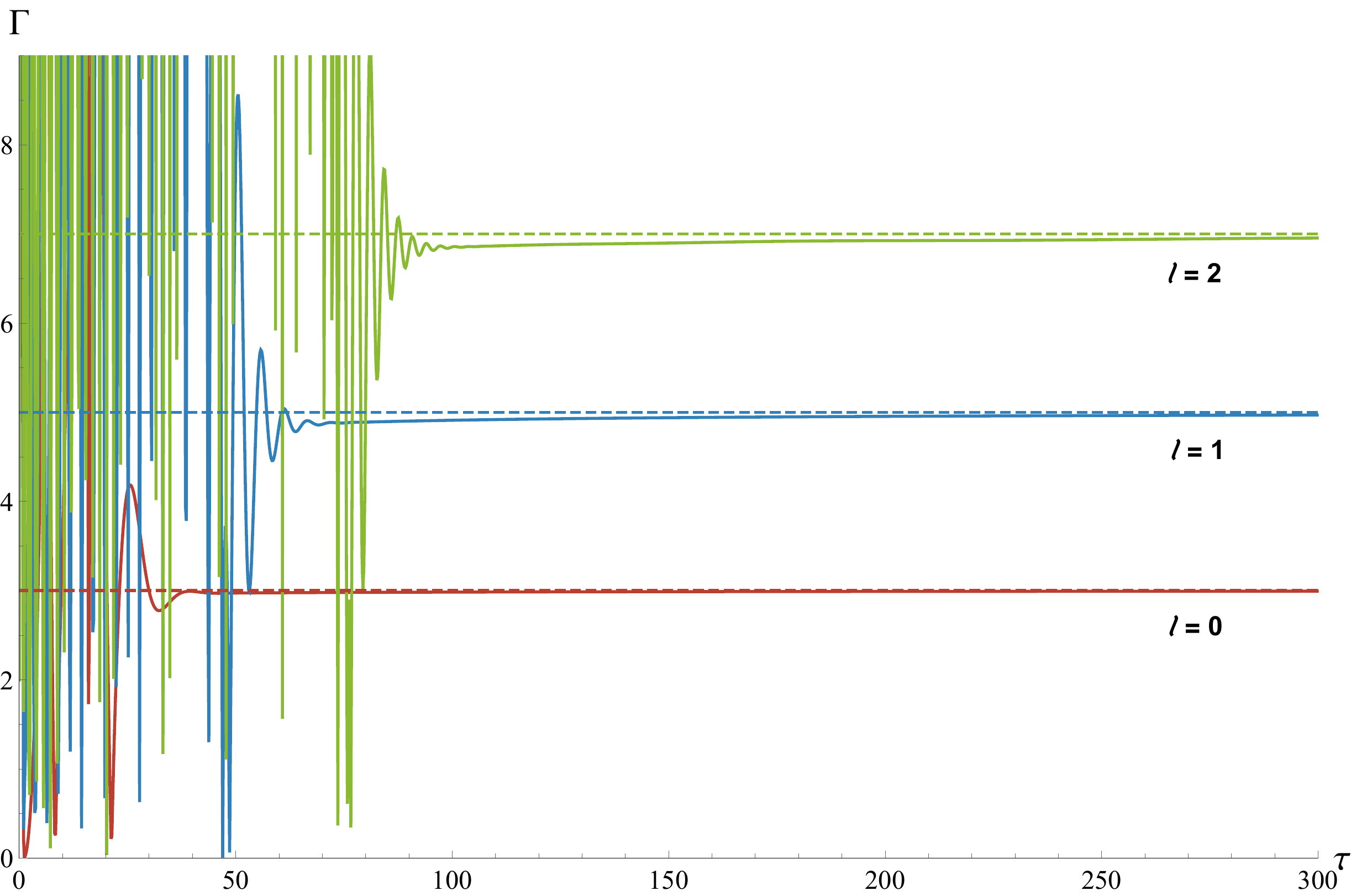}
        \caption{Power law indices for $\sigma>0$}
    \end{subfigure}
    \hfill
    \begin{subfigure}{0.45\textwidth}
        \centering
        \includegraphics[width=\textwidth]{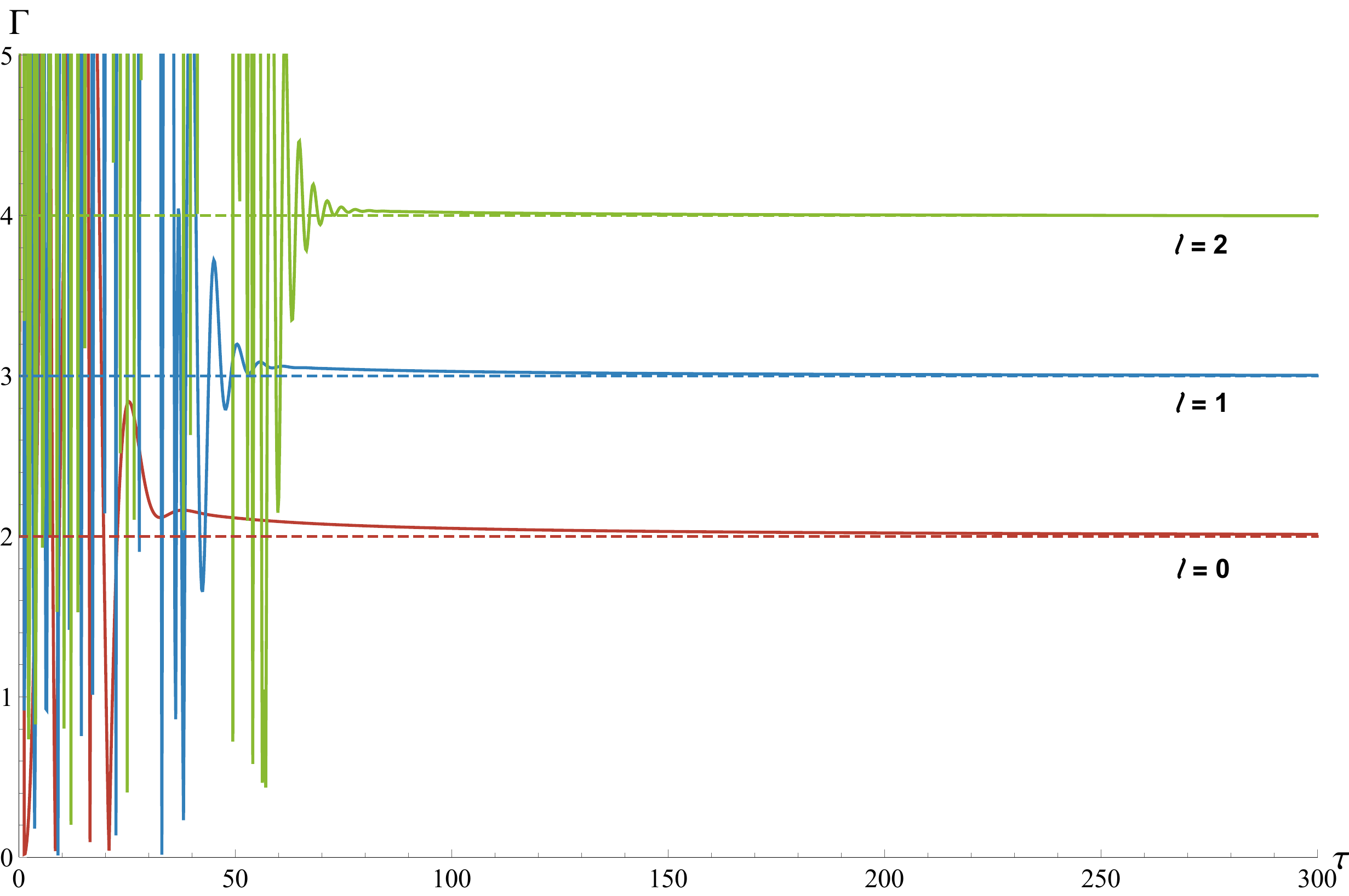}
        \caption{Power law indices for $\sigma=0$}
    \end{subfigure}
    \caption{The effective power law indices as determined by numerical evolution of the multipolar modes $\ell=0$, $\ell=1$, and $\ell=2$. The calculated exponents at late times in the evolutions agree with Price's Law.}
    \label{fig:PriceExponents}
\end{figure}

We validate our code by verifying that the field obeys Price's tail law \cite{PhysRevD.5.2419,PhysRevD.49.883} which states that, at late times, an $\ell-$pole scalar field decays according to
\begin{equation}
    \varphi_\ell\rvert_{\sigma > 0} ~ \sim \tau^{-2 \ell - 3}, \quad  \varphi_\ell|_{\sigma = 0} ~ \sim \tau^{-\ell - 2}.
\end{equation}
We use Gaussian initial data in each $\ell$-mode of $\varphi$,
\begin{equation}
    \varphi_{\ell}(0,\sigma) = \exp \Bigg( -\frac{(\sigma-\sigma_0)^2}{w^2} \Bigg)
\end{equation}
with $\sigma_0 = 0.6$ and $w^2=1/1000$. We use homogeneous data in its derivative, $\partial_\tau \; \varphi_\ell(0,\sigma) = 0$. To quantify how the field decays, we define an \textit{effective power-law index} (following, e.g., \cite{Burko:1997tb,burko_linearized_2018}) for each mode by
\begin{equation}
    \Gamma_\ell = | \tau  \; \partial_\tau \ln \varphi_\ell  |= \bigg\lvert \frac{\tau ~ \Pi_\ell}{\varphi_\ell} \bigg\rvert.
\end{equation}
Observe that, if $\varphi_\ell$ is polynomial in $\tau$, $\Gamma_\ell$ evaluates to the exponent of $\tau$. We perform a simulation with the H2 method, evaluating $\Gamma_\ell$ on $\mathscr{I}^+$ ($\sigma=0$) and a finite distance from $\mathscr{H^+}$ ($\sigma>0$) for the first three harmonic modes ($\ell=0$, $\ell=1$, and $\ell=2$). As shown in Figure \ref{fig:PriceExponents}, $\Gamma_\ell$ converges a constant value corresponding to the correct power-law index for each mode. (For higher order modes, the field decays so rapidly that round-off error obscures the power-law tail. It would be necessary to use extended precision in evolution or use a spatial discretization less prone to round-off error to extract the tails for these modes. In this case, methods such as those based on the Ozaki scheme \cite{10.1145/3472456.3472493} can be used to accelerate DGEMM operations with extended precision on CPU and GPU architectures, but this is beyond the purposes of the present work.)

\begin{figure}
    \centering
    \begin{subfigure}{0.45\textwidth}
        \centering
         \includegraphics[width=\textwidth]{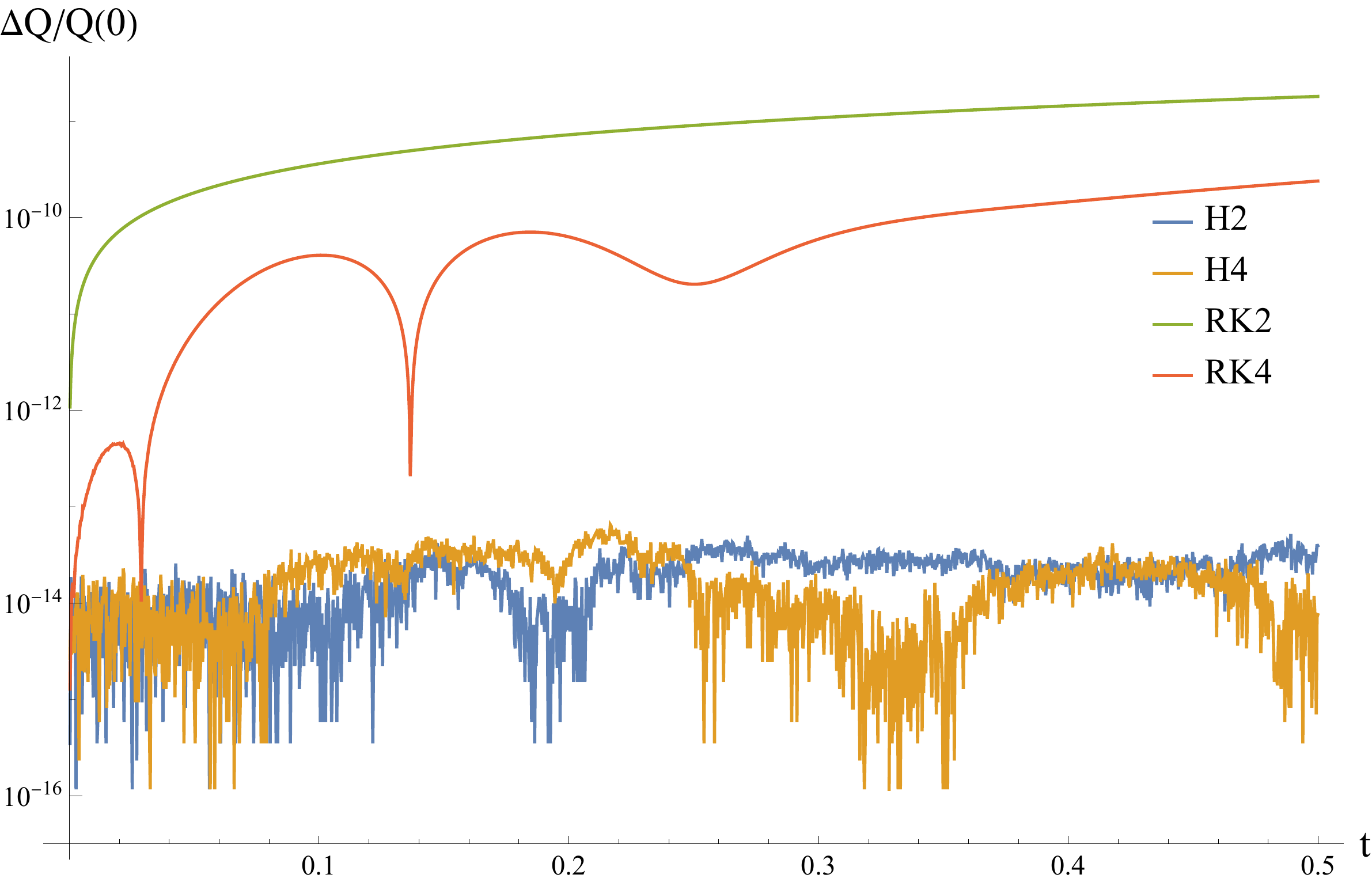}
        \caption{Error in U(1) Gauge Charge}
    \end{subfigure}
    \hfill
    \begin{subfigure}{0.45\textwidth}
        \centering
        \includegraphics[width=\textwidth]{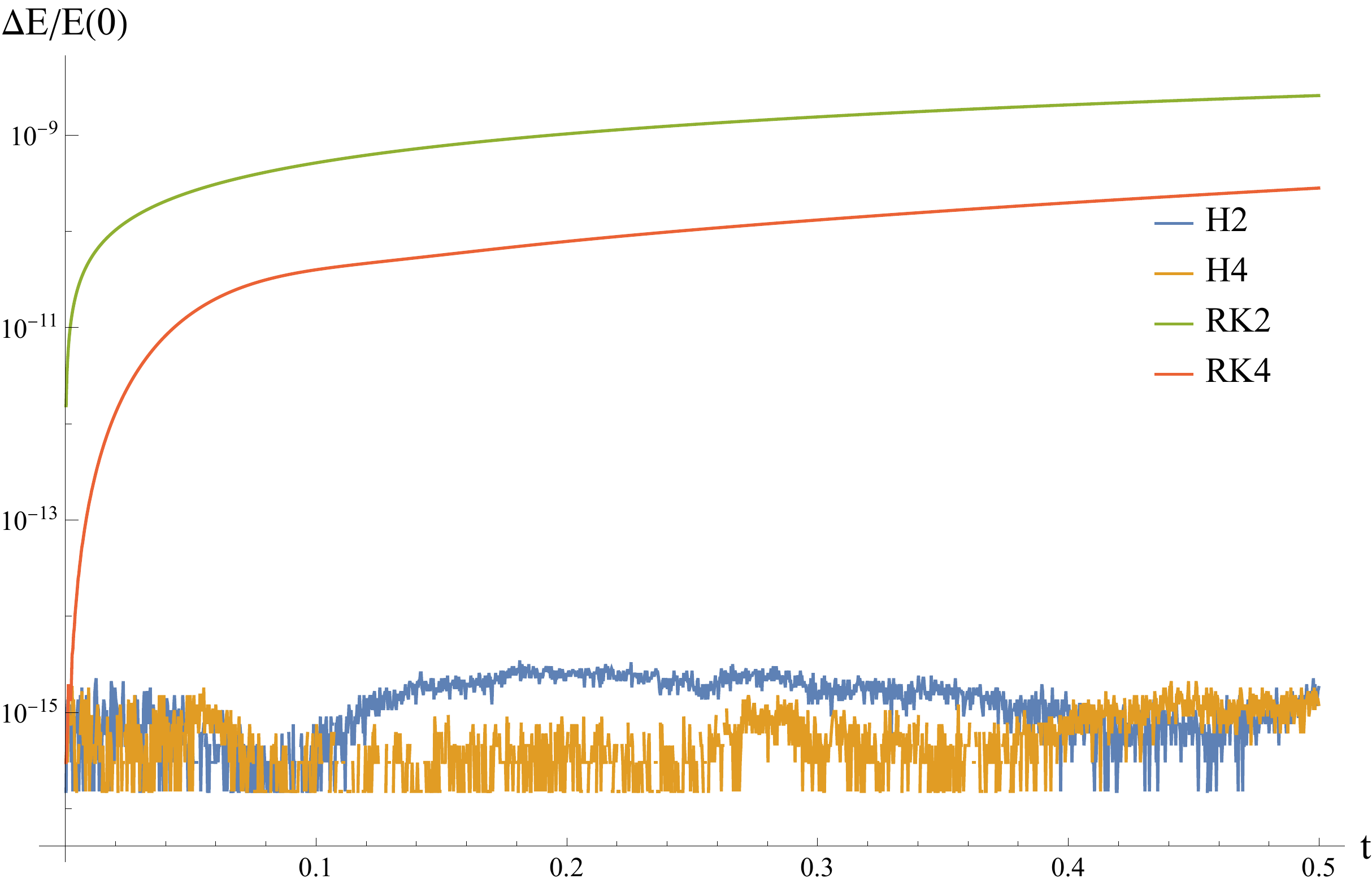}
        \caption{Error in Energy}
    \end{subfigure}
    \caption{The relative error in the conserved U(1) charge $Q$ and energy $E$ for the Klein-Gordon field evolved in the Schwarzschild spacetime. Hermite methods (H2, H4) conserve these quantities to near-machine precision (up to round-off error) while Runge-Kutta methods (RK2, RK4) do not (due to truncation error).}
    \label{fig:KGConserve}
\end{figure}

Having validated our code, we investigate the conservation of Noether charges affiliated with this field. We provide the new initial data
\begin{equation}\label{eq:ComplexData}
    \varphi_{\ell}(0,\sigma) = \exp \Bigg( -\frac{(\sigma-\sigma_0)^2}{w^2} \Bigg) + {\rm{i}} ~\exp \Bigg( -\frac{(\sigma-\sigma_1)^2}{w^2} \Bigg)
\end{equation}
with $\sigma_0 = 0.65$, $\sigma_1 = 0.55$, and $w^2=1/1000$. (The addition of an imaginary part ensures the U(1) charge $Q$ is nonzero.) As before, $\partial_\tau \psi_\ell(0,\sigma) = 0$. We evolve this initial data with a time step of $\Delta \tau = 10^{-5}$ from $\tau=0$ to $\tau=0.5$; this was done so the field with numerically compact support never reached the domain boundaries, which would require the inclusion of boundary flux integrals in the conservation statements (see Appendix \ref{appendix:flux} for a discussion). As we show in Figure \ref{fig:KGConserve}, Hermite methods conserve Eqs.~\eqref{eq:KGGauge} and \eqref{eq:KGEnergy} to machine precision, while explicit Runge-Kutta methods do not.

\subsection{Curvature Perturbations via Bardeen-Press-Teukolsky Functions}
Gravitational perturbations to a non-rotating black hole are typically treated with one of two formalisms: RWZ quantities related to perturbations to the metric tensor, or Bardeen-Press-Teukolsky (BPT) functions related to perturbations to the curvature tensors. Since the RWZ quantities can be treated using largely the same formalism as the Klein-Gordon field (see Appendix \ref{sec:RWZ} for a discussion and a presentation of conserved quantities for these fields), we turn to the BPT formalism in this section.

Instead of scalar fields, BPT quantities $\psi^{(s)}$ are spin-weighted functions \cite{newman_approach_1962}\footnote{That is, they transform as $\psi^{(s)} \rightarrow e^{i s \vartheta}~ \psi^{(s)}$ under frame rotations by an angle $\vartheta$ in the plane orthogonal to the radial direction.}.
Bardeen and Press have shown that such quantities obey a master wave equation in  Schwarzschild spacetime \cite{press_perturbations_1973,doi:10.1063/1.1666175} (their work was extended to rotating black holes by Press and Teukolsky \cite{teukolsky_perturbations_1973,teukolsky_perturbations_1974}). Bini et al. showed that the Bardeen-Press-Teukolsky equation can be written in the covariant form:
\begin{equation}
    g^{\mu \nu}(\nabla_\mu + s \Gamma_\mu)(\nabla_\nu + s \Gamma_\nu) \psi^{(s)} - 4 s^2 \Psi_2 \psi^{(s)} = 0
\end{equation}
where $\nabla_\mu$ is the covariant derivative compatible with the spacetime 4-metric $g_{\mu \nu}$, $\Gamma^\mu$ is a quantity termed the ``connection vector'' and $\Psi_2$ is the non-vanishing Weyl scalar for the unperturbed Type-D black hole spacetime \cite{Bini:2002jx,toth_noether_2018}. Setting $s=0$ would recover the Klein-Gordon field and $s=\pm 1$ would describe electromagnetic test fields. However, we are most interested in gravitational perturbations, described by $s=\pm 2$. This covariant form allows the equations to be cast in alternative coordinate charts, such as hyperboloidal slices \cite{Zenginoglu:2008wc}.

To achieve a $1+1$ formulation in the hyperboloidal coordinates of the previous section, it is necessary to decompose the field into spin-weighted spherical harmonics. Additionally, the equation is singular at both $\sigma=0$ and $\sigma=1$. The quantity $\varphi^{(s)} = (1-\sigma)^s \sigma^{-1-2s} \psi^{(s)}$ can be shown to be regular at both endpoints \cite{zenginoglu_geometric_2011,barack_time-domain_2017}, so we take this as the evolution variable. That is, we perform the decomposition:
\begin{equation}
    \psi^{(s)}(\tau, \sigma, \theta, \phi) = \sum_{\ell = |s|}^\infty ~ \sum_{m = - \ell}^\ell \sigma^{1+2s} (1-\sigma)^{-s}~ \varphi^{(s)}_{\ell m}(\tau, \sigma)~ _s Y_{\ell m}(\theta, \phi).
\end{equation}
Then, the evolution equation for $\varphi^{(s)}_{\ell m}$, introduced in \cite{ansorg_spectral_2016}, takes the regular 1+1 form:
\begin{multline}\label{eq:BPTEffective}
    -(1+\sigma) \partial_\tau^2 \varphi^{(s)} + (1-2\sigma^2) \partial_\tau \partial_\sigma \varphi^{(s)} + (1-\sigma)\sigma^2 \partial_\sigma^2 \varphi^{(s)}+ \sigma ( 2 - 3\sigma + s(2-\sigma) ) \partial_\sigma \varphi^{(s)} \\
    - (2\sigma -s (1-\sigma)) \partial_\tau \varphi^{(s)} - (\ell(\ell+1) +(\sigma-s)(1+s) )\varphi^{(s)} = 0.
\end{multline}
Note that the above equation reduces to Eq.~\eqref{eq:LittlePsiEqn}
for scalar ($s=0$) perturbations, as expected. As in the previous section, we may write this equation in 1+1 covariant form
\begin{equation}
   \eta^{\alpha \beta}(\nabla_{\alpha} +s\Gamma_\alpha)( \nabla_{\beta} +s\Gamma_\beta)\varphi^{(s)} - V_{\ell}^{(s)} \varphi^{(s)} =0
\end{equation}
where the Minkowski 2-metric $\eta_{\alpha \beta}$  is given by Eq.~\eqref{eq:hyperbMink}, $\nabla_\alpha$ is the covariant derivative compatible with $\eta_{\alpha \beta}$, the connection vector has components $\Gamma_\tau=\sigma(2-3\sigma)$ and $\Gamma_\sigma=1+3\sigma$, and the effective potential is given by
$V_\ell^{(s)} = 4 \sigma^2(1-\sigma)(\ell(\ell+1) + (s-\sigma)(1+s))$. We may thus write down a phenomenological action:
\begin{equation}
    S[\varphi^{(s)},\varphi^{(-s)}] = \int_{\mathcal{N}^2} d^2x \sqrt{-\eta} \Big[ \eta^{\alpha \beta}(\nabla_{\alpha} -s\Gamma_\alpha)\varphi^{(-s)}( \nabla_{\beta} +s\Gamma_\beta)\varphi^{(s)} - V_{\ell}^{(s)} \varphi^{(-s)} \varphi^{(s)} \Big].
\end{equation}
Eq.~\eqref{eq:BPTEffective} follows from extremizing the action with respect to $\varphi^{(-s)}$. A conjugate equation for the opposite spin field, $\varphi^{(-s)}$, follows from extremizing the action with respect to $\varphi^{(s)}$. This conjugate equation can be obtained by changing $s\rightarrow-s$ in Eq.~\eqref{eq:BPTEffective}.

The methods of the previous section are immediately applicable and we write down a conserved U(1) gauge charge,
\begin{multline}
    Q = \int_{\Sigma_\tau} d\sigma \bigg[ (1+\sigma) (\varphi^{(-s)} \partial_\tau \varphi^{(s)} - \varphi^{(s)} \partial_\tau \varphi^{(-s)} )\\
    - \frac{1-2\sigma^2}{2} (\varphi^{(-s)} \partial_\sigma \varphi^{(s)} - \varphi^{(s)} \partial_\sigma \varphi^{(-s)} ) - s (1-\sigma) \varphi^{(-s)} \varphi^{(s)} \bigg]
\end{multline}
and a conserved energy
\begin{multline}
    E = \int_{\Sigma_\tau} d\sigma \bigg[ (1+\sigma) \partial_\tau \varphi^{(-s)} \partial_\tau \varphi^{(s)} + \sigma^2 (1-\sigma) \partial_\sigma \varphi^{(-s)} \partial_\sigma \varphi^{(s)} \\
    - \frac{s\sigma(2-\sigma)}{2} (\varphi^{(-s)} \partial_\sigma \varphi^{(s)} - \varphi^{(s)} \partial_\sigma \varphi^{(-s)}) + (\ell(\ell+1) - s^2 + \sigma) \varphi^{(-s)} \varphi^{(s)} \bigg]
\end{multline}
(cf. Ref.~\cite{toth_noether_2018} for a covariant derivation of these Noether charges). Since $Q$ and $E$ involve two fields of opposite spin weights, it is necessary to evolve both fields of opposite spin at once. 

We numerically study this problem by examining U(1) charge and energy conservation for a gravitational perturbation ($|s|=2$) in the quadrupolar mode ($\ell=2$). We study it under the same conditions as the scalar field of the previous section. We use the initial data of Eq.~\eqref{eq:ComplexData} in the $s=2$ field and its complex conjugate in the $s=-2$ field. Unlike with the previous problems, we find that the Noether charges are not exactly conserved with Hermite methods. We do, however, note that the error is bounded, whereas it grows without bound when explicit Runge-Kutta methods are employed.
\begin{figure}
    \centering
    \begin{subfigure}{0.45\textwidth}
        \centering
        \includegraphics[width=\textwidth]{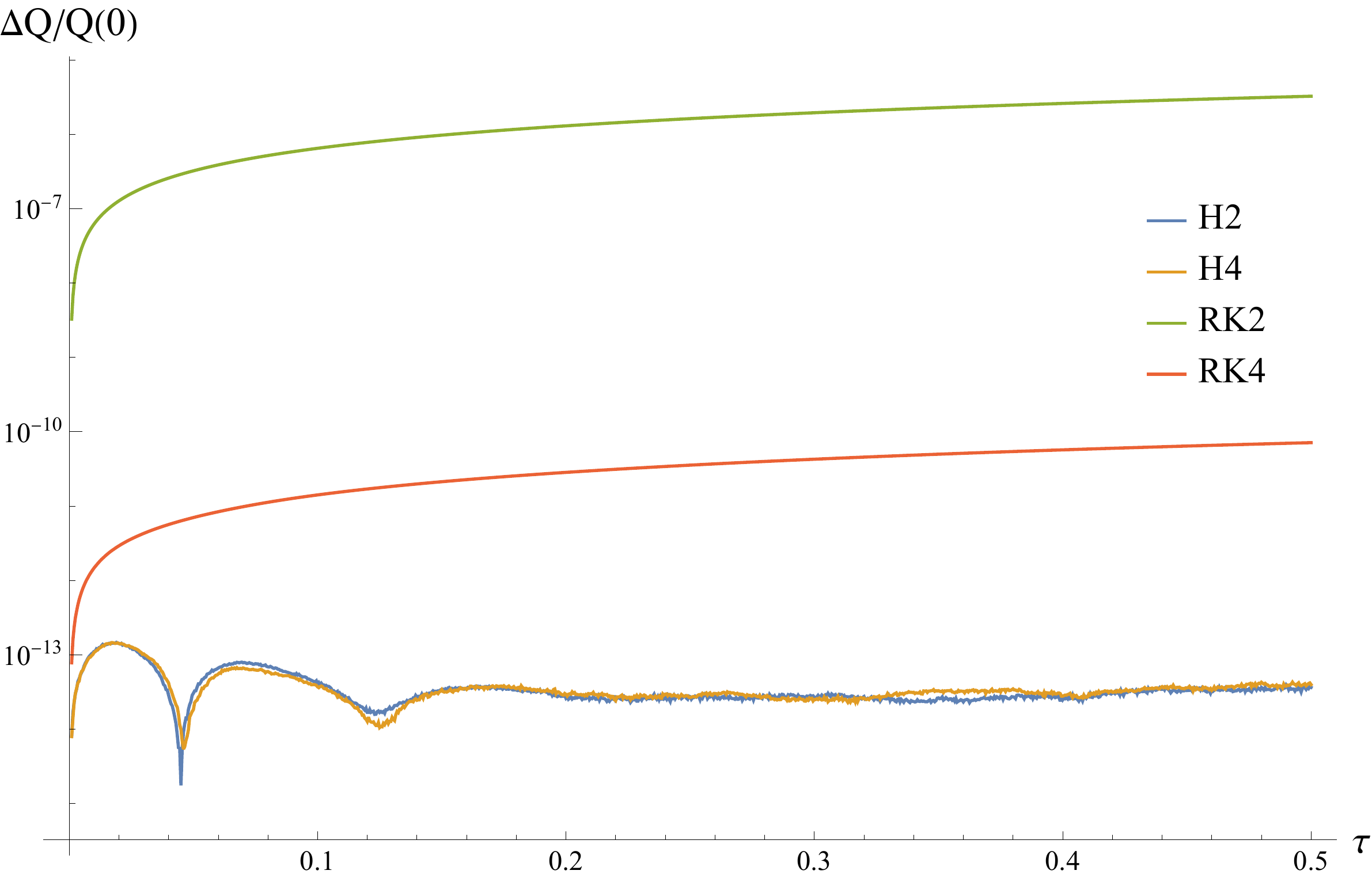}
        \caption{Error in U(1) Gauge Charge}
    \end{subfigure}
    \hfill
    \begin{subfigure}{0.45\textwidth}
        \centering
        \includegraphics[width=\textwidth]{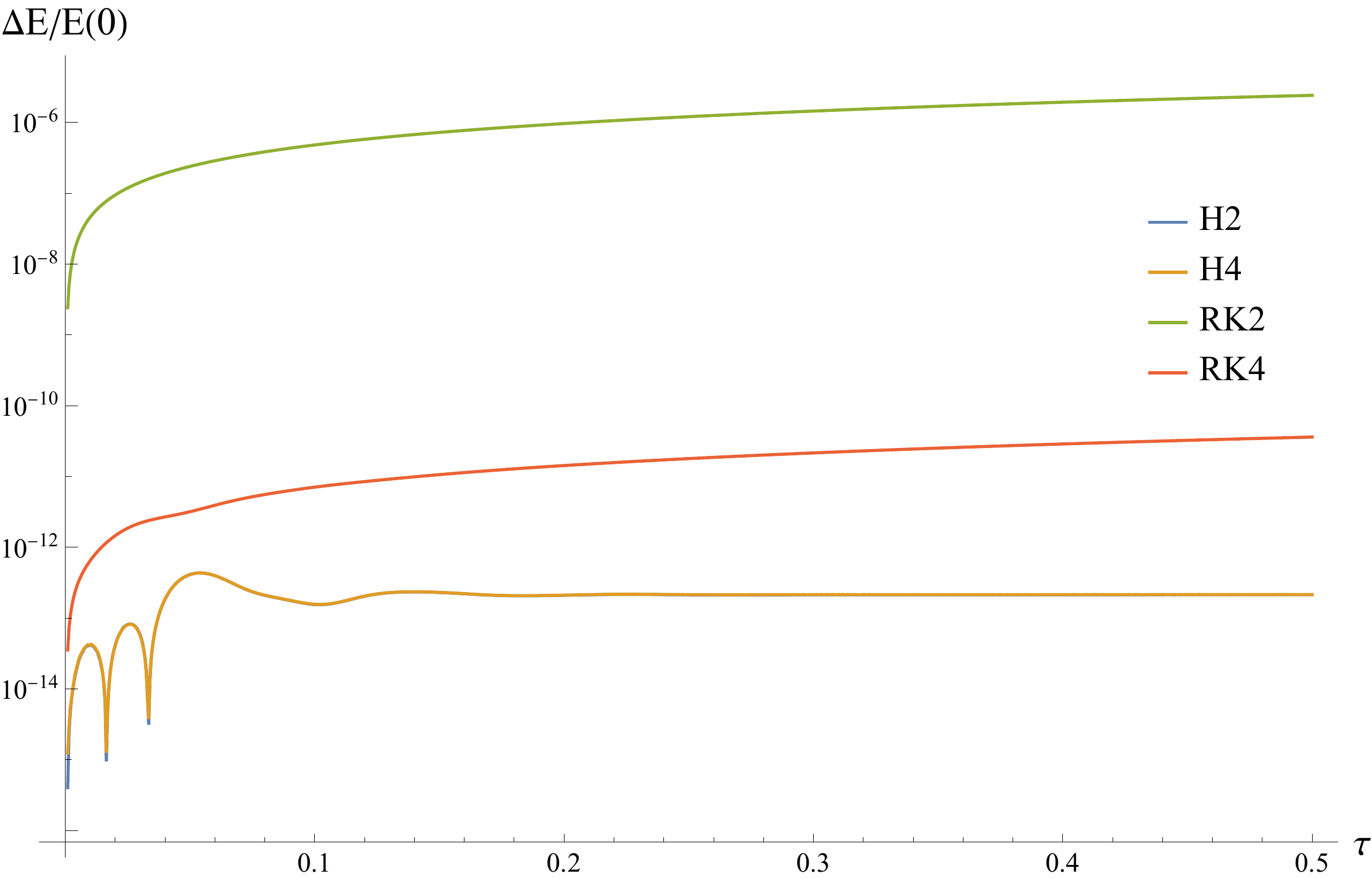}
        \caption{Error in Energy}
    \end{subfigure}
    \caption{The relative error in the conserved U(1) charge $Q$ and energy $E$ for the $|s|=2$ BPT equations. Although neither quantity is exactly conserved, we note that the error with Hermite methods is bounded, while it grows without bound with Runge-Kutta methods.}
    \label{fig:BPTConserve}
\end{figure}

\section{Summary and Outlook}
The pressing need for accurate EMRI waveform models within the next decade has brought the importance of accurate numerical simulation to the forefront of relativistic astrophysics. In particular, the need to evolve for timescales much larger than past numerical relativity simulations in a manner which accurately tracks the quantities of importance in GW astronomy, energy and angular momentum, brings forth the need to adopt numerical methods suited to the simulation of physical problems.

Standard explicit Runge-Kutta methods, while incredibly popular and simple to implement (especially for linear equations), are ill-suited to the EMRI problem. They possess CFL limits, severely limiting the size of the time step which can be used and therefore require many floating-point operations to perform a long time evolution. In addition, these methods do not respect the qualitative features of classical dynamics: time-reversal symmetry and symplecticity. Therefore, a numerical result obtained using such methods should not be expected to accurately reflect the physics of the problem in question.

In this work, we have demonstrated that a class of implicit methods, generalized Hermite integration schemes, which are CFL unlimited and are automatically symmetric under time reversal. We have shown that they are also volume-preserving in phase space for linear problems and that they numerically conserve  Noether charges of several field theories. We have formulated phenomenological actions for the fields of BHPT and applied Noether's theorem to derive constants of evolution common to all of them: a charge corresponding to U(1) gauge symmetry in the field and an energy corresponding to the stationarity of the spacetime.

The Hermite integration methods outlined in this paper have also been shown to work when a point-particle source term is added to the flat spacetime Klein-Gordon equation. In this case, the Hermite integration methods must be modified to accommodate discontinuous functions, which we have shown in \cite{Markakis:2014nja} for the second order method. (The appropriate generalization for the fourth order method will be presented in a subsequent paper). This indicates that the methods presented here will be suitable for the problem of gravitational perturbations sourced by a point mass, the theoretical underpinning of EMRI physics. We will explore this application in subsequent work. 

In addition, although the methods outlined here were demonstrated for linearized PDEs arising in BH perturbation theory, the methods are applicable to the non-linear PDEs of full numerical general relativity as well. This leads to implicit time-symmetric schemes that have a linear and a non-linear part. In this case, one can treat the linear part of the scheme explicitly (by matrix inversion) and the non-linear part implicitly (by self-consistent iteration). The conservation and stability properties of such schemes will be demonstrated in future work.

The methods outlined here also pave the way for evolving perturbations on a rotating (Kerr) black hole spacetime background. This case has the added complication that, upon spin-weighted spherical harmonic decompostion, $m$-modes are decoupled (due to axisymmetry), but $\ell$-modes are coupled to next to nearest neighbors in 1+1 dimensions (due to lack of spherical symmetry)\cite{barack_time-domain_2017}. While this complication means that large mode-coupling matrices have to be inverted, once the coupled system of all $\ell$-modes is written in the first-order form of Eq.~\eqref{eq:FirstOrderReduced}, and the dicretized matrix operator $\bf{L}$ is evaluated on a grid and stored, the time-symmetric integration schemes outlined in Sec.~\ref{Hermite} are readily and easily applicable.

Finally, from a numerical computing perspective, it has been demonstrated that time integration via symmetric methods is highly accurate for evolving the wave-type PDEs of black hole perturbation theory, and the residual error is mainly due to spatial discretization. Pseudospectral methods converge rapidly and are suitable for spatial discretization, but general matrix multiplication libraries on CPUs and GPUs prioritize performance and parallelization rather than accuracy. As a result, the vast majority of libraries do not use compensated summation, and this causes round-off error to accumulate. The development and use of general matrix multiplication libraries that use compensated summation is necessary in order to avoid the accumulation of round-off error from spatial differentiation.

\section*{Acknowledgments}
We thank Derek Glennon for his great assistance in testing Hermite methods for the advection equation, Abhay Shah for helpful discussions on the BPT equation, and An{\i}l Zengino\u{g}lu for valuable comments and suggestions on the derivation of hyperboloidal slices. C.M. was supported by the European Union’s Horizon 2020 research and innovation programme under the Marie Skłodowska-Curie grant agreement
No 753115.

\appendix
\section{Modified Hamilton Equations}\label{sec:NewHamilton}
The Hamiltonian density describing the Schr\"{o}dinger field (Eq.~\eqref{eq:SchroHamiltonian}) is unusual in that it contains explicit dependence on spatial derivatives of the canonical momentum. It is thus necessary to modify the standard Hamilton equations of motion.

We begin by considering an action functional
\begin{align}
    S &= \int dt dx \mathcal{L}\\
    & = \int dt dx ( \pi \partial_t \psi - \mathcal{H} )
\end{align}
Although we are only considering one field here, the generalization to multiple fields is obvious. When we vary with respect to the quantities $\psi$ and $\pi$, we note that $\mathcal{H} = \mathcal{H}(\psi, \partial_x \psi, \pi, \partial_x \pi )$, so
\begin{equation}
    \delta S = \int dt dx \bigg( \pi \partial_t \delta \psi + \delta \pi \partial_t \psi - \frac{\partial \mathcal{H}}{\partial \psi} \delta \psi - \frac{\partial \mathcal{H}}{\partial (\partial_x \psi)} \partial_x \delta \psi - \frac{\partial \mathcal{H}}{\partial \pi} \delta \pi - \frac{\partial \mathcal{H}}{\partial (\partial_x \pi)} \partial_x \delta \pi \bigg)
\end{equation}
After integrating by parts and demanding that $\delta \psi$ and $\delta \pi$ vanish on the domain boundaries, we are left two requirements for $\delta S =0$:
\begin{equation}
    \partial_t \psi = \frac{\partial \mathcal{H}}{\partial \pi} - \partial_x \bigg( \frac{\partial \mathcal{H}}{\partial (\partial_x \pi)} \bigg)
\end{equation}
\begin{equation}
    \partial_t \pi = - \frac{\partial \mathcal{H}}{\partial \psi} + \partial_x \bigg( \frac{\partial \mathcal{H}}{\partial (\partial_x \psi)} \bigg)
\end{equation}

\section{Evolution Schemes for Second-Order Equations}\label{sec:ExplicitSchemes}
Here, we present an alternative to the method for second-order-in-time PDEs discussed in Section \ref{sec:SecondOrder}. Rather than define the new ``state vector'' $V$ combining  $\Psi$ and $\Pi$, we separately apply Hermite integration rules to the evolution equations for these quantities. For compactness, we introduce the new notation
\begin{equation}
    \partial_t \Psi = \Pi
\end{equation}
\begin{equation}
    \partial_t \Pi = L_{2 2} \Pi + L_{2 1} \Psi
\end{equation}
where $L_{21}$ and $L_{22}$ are the differential operators appearing in Eq.~\eqref{eq:Generic1p1} (and the bottom rows of the matrix $L$). Applying Eq.~\eqref{eq:LD2}, we find that, at $\mathcal{O}(\Delta t^2)$, it is possible to obtain an explicit expression for $\Pi^{n+1}$:
\begin{equation}\label{eq:LD2PiScheme}
    \Pi^{n+1} = \bigg( I - \frac{\Delta t}{2} L_{2 2} - \frac{\Delta t^2}{4}
L_{2 1} \bigg)^{-1} \Bigg[ \bigg( I + \frac{\Delta t}{2} L_{2 2} + \frac{\Delta
t^2}{4} L_{2 1} \bigg) \Pi^n + \Delta t L_{2 1} \Psi^n \Bigg]
\end{equation}
This may now be directly substituted into
\begin{equation}
    \Psi^{n+1} = \Psi^n + \frac{\Delta t}{2}(\Pi^n +\Pi^{n+1})
\end{equation}
to find $\Psi^{n+1}$. When Eq.~\eqref{eq:LD4} is applied, we find, at $\mathcal{O}(\Delta t^4)$, an explicit expression for $\Pi^{n+1}$,
\begin{multline}
    \Pi^{n+1} = \Bigg[ I - \frac{\Delta t}{2} L_{2 2} + \frac{\Delta t^2}{12} \bigg( L_{2 2}^2 + L_{2 1} \bigg) - \bigg( I - \frac{\Delta t}{6} L_{2 2} \bigg) \bigg(\frac{\Delta t^2}{4} L_{2 1} \bigg) \bigg( I + \frac{\Delta t^2}{12} L_{2 1} \bigg)^{-1}\\
    + \bigg( I - \frac{\Delta t}{6} L_{2 2} \bigg) \bigg(\frac{\Delta t^2}{4} L_{2 1} \bigg) \bigg( I + \frac{\Delta t^2}{12} L_{2 1} \bigg)^{-1} \bigg( \frac{\Delta t}{6} L_{2 2} \bigg) \Bigg]^{-1}\\
    \Bigg\{ \Bigg[ I + \frac{\Delta t}{2} L_{2 2} + \frac{\Delta t^2}{12} \bigg( L_{2 2}^2 + L_{2 1} \bigg) + \bigg( I + \frac{\Delta t}{6} L_{2 2} \bigg) \bigg(\frac{\Delta t^2}{4} L_{2 1} \bigg) \bigg( I + \frac{\Delta t^2}{12} L_{2 1} \bigg)^{-1}\\
    + \bigg( I - \frac{\Delta t}{6} L_{2 2} \bigg) \bigg(\frac{\Delta t^2}{4} L_{2 1} \bigg) \bigg( I + \frac{\Delta t^2}{12} L_{2 1} \bigg)^{-1} \bigg( \frac{\Delta t}{6} L_{2 2} \bigg) \Bigg] \Pi^n + \Delta t L_{2 1} \Psi^n \Bigg\}.
\end{multline}
This may now be directly inserted into 
\begin{equation}
    \Psi^{n+1} = \Psi^n + \frac{\Delta t}{2} L_{2 2} (\Pi^n + \Pi^{n+1}) + \frac{\Delta t^2}{12}[  L_{2 1} (\Psi^n + \Psi^{n+1}) + L_{2 2} (\Pi^n - \Pi^{n+1}) ]
\end{equation}
to find the new field  $\Psi$, which can be obtained by algebraically solving the above equation
for $\Psi^{n+1}$.

\section{Boundary Flux}\label{appendix:flux}
The derivations of gauge charge and energy conservation in the main text presuppose that the field remains within the computational domain at all times. However, the astrophysically observable quantity  is usually not the field itself by the radiation it creates. Modelling this amounts to allowing the field to irreversibly exit the domain by an appropriate choice of boundary conditions (automatically enforced with hyperboloidal compactification; see the main text).

If the field is allowed to irreversibly radiate, then energy and other quantities are no longer conserved. What the Noether conservation laws instead yield are \textit{flux-balancing statements}: the rate at which an integral charge changes over the domain must be equal to the currents at the boundaries. To make this statement mathematically precise, suppose there is a conserved Noether current $J^\mu$ (following from spacetime translation symmetry or gauge symmetry, e.g.). The local conservation law reads
\begin{equation}
    \nabla_\mu J^\mu = \frac{1}{\sqrt{-g}} \partial_\mu \Big( \sqrt{-g}~ J^\mu \Big) = 0
\end{equation}
Assuming 1+1 spacetime dimensions, we may obtain a global conservation law becomes by integrating this equation over a complete time slice:
\begin{equation}
    \int_{\Sigma_0} dx^1~ \partial_0 \Big( \sqrt{-g}~ J^0 \Big) = -  \int_{\Sigma_0} dx^1~ \partial_1 \Big(\sqrt{-g}~ J^1 \Big)
\end{equation}
We may apply the fundamental theorem of calculus to the right hand side, and we may note that the region of integration does not depend on $x^0$. So, we find that
\begin{equation}
    \frac{d}{d x^0} \int_{\Sigma_0} dx^1 \sqrt{-g}~ J^0 = -  \sqrt{-g}~ J^1 \Big |_{\partial \Sigma_0}
\end{equation}
The integral on the left hand side matches the definition of a global charge in Eq.~\eqref{eq:GenericCharge}. However, we now note that its time derivative does not vanish if $J^1$ is nonzero at the boundary of the time slice.

\begin{figure}
    \centering
    \includegraphics[width=0.5\textwidth]{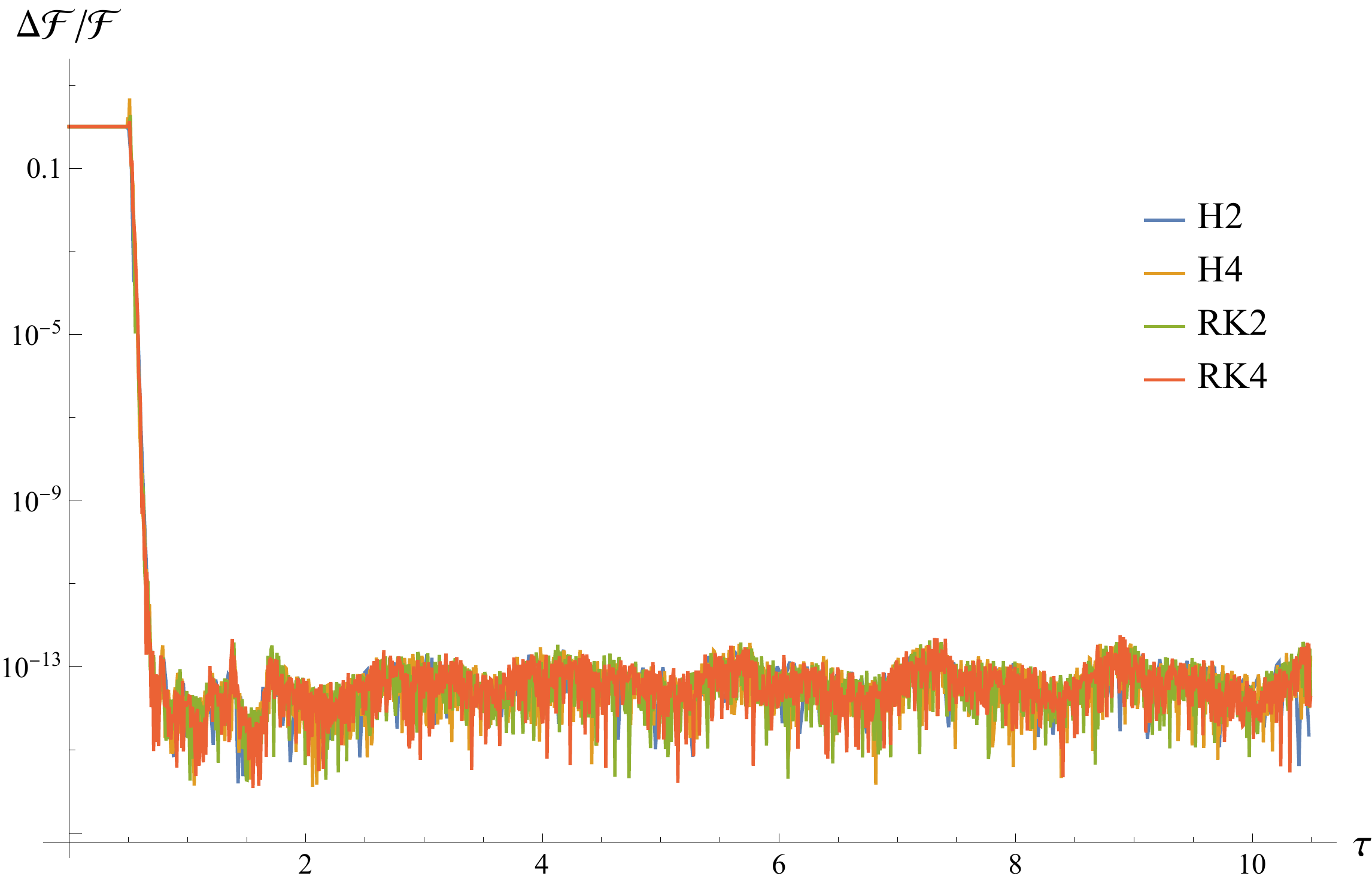}
    \caption{The relative error between $d E / d\tau$ and $\mathcal{F}(\partial \Sigma_0)$ for H and RK methods. The value of unity at the start of the simulation is attributed to the fact that the initial data has effectively compact support rendering $\mathcal{F}(\partial \Sigma_0) = 0$ while round-off errors prevent the domain-wide integral $d E / d\tau$ from vanishing.}
    \label{fig:Flux}
\end{figure}

Take the Klein-Gordon field in Schwarzschild spacetime as an example. Using the effective action~\eqref{eq:ActionEffective}, giving rise to the energy~\eqref{eq:KGEnergy}, the conservation statement becomes
\begin{equation}
    \frac{d E}{d \tau} = \mathcal{F}(\partial \Sigma_0) = -\partial_\tau \varphi^\star \partial_\tau \varphi|_{\sigma = 1} - \partial_\tau \varphi^\star \partial_\tau \varphi|_{\sigma = 0}
\end{equation}
If Hermite integration schemes or other time-symmetric methods are to be useful in gravitational wave and self-force calculations, they must accurately track the change in the field's global Noether charges as the system evolves. We evaluate the accuracy of such methods by considering the relative error between $\mathcal{F}(\partial \Sigma_0)$ and $d E / d\tau$. We note that the $\tau$ derivative of $E$ may be evaluated by moving the derivative inside the integral and invoking the evolution equation to remove any $\partial^2_\tau \varphi$ terms. The results are shown in Figure \ref{fig:Flux}. We note that both H and RK methods accurately track the loss of energy.

\section{Noether charges for the RWZ Equations}\label{sec:RWZ}
Martel and Poisson present a formalism describing metric (spin $s=2$) perturbations of non-rotating black holes that is not only gauge-independent but fully covariant on $\mathcal{M}^2$ \cite{martel_gravitational_2005}. They show it can be described by two scalar quantities: the Zerilli-Moncrief function for even-parity perturbations \cite{PhysRevD.2.2141,PhysRevD.55.2124} and the gauge-invariant Cunningham-Price-Moncrief \cite{1978ApJ} function (rather than the classic, gauge-dependent, Regge-Wheeler function \cite{PhysRev.108.1063}) for odd parity perturbations. When the Schwarzschild spacetime is decomposed into $\mathcal{M}^2 \times S^2$, as in Section \ref{sec:SecondOrder}, each of these two functions is described by a master wave equation with a potential term:
\begin{equation}
    \Box \Psi - V_\ell \Psi = S_{\ell m}
\end{equation}
where $S_{\ell m}$ is a source term derived from the stress-energy tensor of the matter projected onto spherical harmonics \cite{martel_gravitational_2005}.

The potential $V_\ell$ depends on the parity of the perturbation. For odd perturbations, it reads (in the hyperboloidal coordinates of Sec. \ref{sec:BHScalar})
\begin{equation}
    V_\ell^{(\text{o})} =
    4 \sigma^2(1-\sigma)(\ell(\ell+1) -3\sigma)
\end{equation}
For even perturbations, it reads
\begin{equation}
    V_\ell^{(\text{e})}  = 4 \sigma^2(1-\sigma)\frac{9 \sigma^2 (\mu + \sigma) + \mu (\ell(\ell+1) + 3 \sigma)}{\mu + 3 \sigma}
\end{equation}
where $\mu = (\ell-1)(\ell+2)$ (note that valid perturbations are only described by $\ell \geq 2$ \cite{martel_gravitational_2005}).

With these expressions for the potential, and if we set the source term to zero to describe vacuum perturbations, the conserved quantities derived for the Klein-Gordon equation have direct analogs in this problem. Starting from the odd-sector equation,
\begin{equation}
    \Box \Psi_{\mathrm{o}} - V^{(\text{o})}_\ell \Psi_{\mathrm{o}} = 0,
\end{equation}
we use the hyperboloidal coordinates introduced before to write this equation as
\begin{multline}
    -(1+\sigma) \partial_\tau^2 \Psi_{\mathrm{o}} + (1 - 2\sigma^2) \partial_\tau \partial_\sigma \Psi_{\mathrm{o}} + \sigma^2 (1-\sigma) \partial_\sigma^2  \Psi_{\mathrm{o}}\\
    - 2 \sigma \partial_\tau \Psi_{\mathrm{o}} + \sigma(2-3\sigma) \partial_\sigma \Psi_{\mathrm{o}} - (\ell(\ell+1) - 3 \sigma) \Psi_{\mathrm{o}} = 0
\end{multline}
Using the effective metric to write a phenomenological action,
\begin{equation}
    S[\Psi_{\mathrm{o}},{\Psi_{\mathrm{o}}}\!\!\!\!\!\!\!\!^{\star}] = \int_{\mathcal{N}^2} d^2 x \sqrt{-\eta} \Big(\eta^{\alpha \beta} \nabla_{\alpha} \Psi_{\mathrm{o}} \nabla_{\beta} {\Psi_{\mathrm{o}}}\!\!\!\!\!\!\!\!^{\star} - V_\ell^{(\text{o})} {\Psi_{\mathrm{o}}}\!\!\!\!\!\!\!\!^{\star} \Psi_{\mathrm{o}} \Big),
\end{equation}
we may write down a conserved U(1) charge
\begin{equation}\label{eq:RWGauge}
    Q = \int_{\Sigma_\tau} d\sigma \bigg( (1+\sigma) ({\Psi_{\mathrm{o}}}\!\!\!\!\!\!\!\!^{\star} \partial_\tau \Psi_{\mathrm{o}} - \Psi_{\mathrm{o}}\partial_\tau {\Psi_{\mathrm{o}}}\!\!\!\!\!\!\!\!^{\star} ) - \frac{1-2\sigma^2}{2} ({\Psi_{\mathrm{o}}}\!\!\!\!\!\!\!\!^{\star} \partial_\sigma \Psi_{\mathrm{o}} - \Psi_{\mathrm{o}} \partial_\sigma {\Psi_{\mathrm{o}}}\!\!\!\!\!\!\!\!^{\star} ) \bigg)
\end{equation}
and an energy
\begin{equation}\label{eq:RWEnergy}
    E = \int_{\Sigma_\tau} d\sigma \bigg( (1+\sigma) \partial_\tau {\Psi_{\mathrm{o}}}\!\!\!\!\!\!\!\!^{\star} \partial_\tau \Psi_{\mathrm{o}} + \sigma^2 (1-\sigma) \partial_\sigma {\Psi_{\mathrm{o}}}\!\!\!\!\!\!\!\!^{\star} \partial_\sigma \Psi_{\mathrm{o}} + (\ell(\ell+1) - 3 \sigma) {\Psi_{\mathrm{o}}}\!\!\!\!\!\!\!\!^{\star} \Psi_{\mathrm{o}} \bigg)
\end{equation}

Similarly, the even sector equation reads:
\begin{equation}
    \Box \Psi_{\mathrm{e}} - V^{(\text{e})}_\ell \Psi_{\mathrm{e}} = 0.
\end{equation}
We use the hyperboloidal coordinates introduced before to write this equation as
\begin{multline}
    -(1+\sigma) \partial_\tau^2 \Psi_{\mathrm{e}} + (1 - 2\sigma^2) \partial_\tau \partial_\sigma \Psi_{\mathrm{e}} + \sigma^2 (1-\sigma) \partial_\sigma^2  \Psi_{\mathrm{e}}\\
    - 2 \sigma \partial_\tau \Psi_{\mathrm{e}} + \sigma(2-3\sigma) \partial_\sigma \Psi_{\mathrm{e}} - \frac{9 \sigma^2 (\mu + \sigma) + \mu (\ell(\ell+1) + 3 \sigma)}{\mu + 3 \sigma} \Psi_{\mathrm{e}} = 0
\end{multline}
Using the effective metric to write a phenomenological action,
\begin{equation}
    S[\Psi_{\mathrm{e}},{\Psi_{\mathrm{e}}}\!\!\!\!\!\!\!\!^{\star}] = \int_{\mathcal{N}^2} d^2 x \sqrt{-\eta} \Big(\eta^{\alpha \beta} \nabla_{\alpha} \Psi_{\mathrm{e}} \nabla_{\beta} {\Psi_{\mathrm{e}}}\!\!\!\!\!\!\!\!^{\star} - V_\ell^{(\text{e})} {\Psi_{\mathrm{e}}}\!\!\!\!\!\!\!\!^{\star} \Psi_{\mathrm{e}} \Big)
\end{equation}
we may write down a conserved U(1) charge
\begin{equation}\label{eq:ZGauge}
    Q = \int_{\Sigma_\tau} d\sigma \bigg( (1+\sigma) ({\Psi_{\mathrm{e}}}\!\!\!\!\!\!\!\!^{\star} \partial_\tau \Psi_{\mathrm{e}} - \Psi_{\mathrm{e}} \partial_\tau {\Psi_{\mathrm{e}}}\!\!\!\!\!\!\!\!^{\star} ) - \frac{1-2\sigma^2}{2} ({\Psi_{\mathrm{e}}}\!\!\!\!\!\!\!\!^{\star} \partial_\sigma \Psi_{\mathrm{e}} - \Psi_{\mathrm{e}} \partial_\sigma {\Psi_{\mathrm{e}}}\!\!\!\!\!\!\!\!^{\star} ) \bigg)
\end{equation}
and a conserved energy
\begin{multline}\label{eq:ZEnergy}
    E = \int_{\Sigma_\tau} d\sigma \Bigg( (1+\sigma) \partial_\tau {\Psi_{\mathrm{e}}}\!\!\!\!\!\!\!\!^{\star} \partial_\tau \Psi_{\mathrm{e}} + \sigma^2 (1-\sigma) \partial_\sigma {\Psi_{\mathrm{e}}}\!\!\!\!\!\!\!\!^{\star} \partial_\sigma \Psi_{\mathrm{e}}\\
    + \frac{9 \sigma^2 (\mu + \sigma) + \mu (\ell(\ell+1) + 3 \sigma)}{\mu + 3 \sigma} {\Psi_{\mathrm{e}}}\!\!\!\!\!\!\!\!^{\star} \Psi_{\mathrm{e}} \Bigg).
\end{multline}

\bibliography{references}

\end{document}